\documentclass{aci}

\usepackage{txfonts}

\usepackage[utf8]{inputenc}
\usepackage{url}
\usepackage{hyperref}
\usepackage{algorithm}
\usepackage{algorithmic}
\usepackage{array}
\usepackage{diffcoeff}
\usepackage{subcaption}

\usepackage{booktabs}
\usepackage{longtable}

\newcommand{\platfont}[1]{\textsc{#1}}

\def\typeofarticle{Research article}
\def\currentvolume{6}
\def\currentissue{1}
\def\currentyear{2026}

\def\ppages{xxx--xxx.}
\def\DOI{10.3934/aci.2026xxx}
\def\Received{20 April 2026}
\def\Revised{17 June 2026}
\def\Accepted{17 June 2026}
\def\Published{ June 2026}
\usepackage{cite}
\begin{document}

\title{Dementia classification from spontaneous speech using wrapper-based feature selection}

\author{%
  Marko Niemelä\affil{1,2,}\corrauth,
  Mikaela von Bonsdorff\affil{2,3},
  Sami Äyrämö\affil{1,4}
  and
  Tommi Kärkkäinen\affil{1}
}

\shortauthors{the Author(s)}

\address{%
  \addr{\affilnum{1}}{Faculty of Information Technology, University of Jyväskylä, Jyväskylä, Finland}
	\addr{\affilnum{2}}{Public Health Programme, Folkhälsan Research Center, Helsinki, Finland}  
  \addr{\affilnum{3}}{Faculty of Sport and Health Sciences and Gerontology Research Center, University of Jyväskylä, Jyväskylä, Finland}
	\addr{\affilnum{4}}{Wellbeing Services County of Central Finland, Finland}
	}

\corraddr{Email: marko.nieme@gmail.com.}

\editor{Pasi Fränti}

\begin{abstract}
Dementia encompasses a group of syndromes that impair cognitive functions such as memory, reasoning, and the ability to perform daily activities. As populations globally age, nearly 10 million new dementia cases occur annually. Clinical diagnosis remains challenging because symptoms overlap with other conditions and require comprehensive cognitive assessment, highlighting the need for feasible and accurate detection methods. Recent advances in machine learning have highlighted spontaneous speech as a promising noninvasive, cost-effective, and scalable biomarker for dementia detection. In this study, spontaneous speech recordings from the ADReSS dataset and the extended Pitt Corpus were analyzed, consisting of picture description tasks performed by cognitively healthy individuals and participants with Alzheimer's disease or dementia. Unlike many prior approaches relying on speech-active segments, acoustic features were extracted from entire recordings with the openSMILE toolkit. This recording-level representation reduces the number of feature vectors and provides a computationally efficient framework for dementia classification, while indirectly incorporating pause- and hesitation-related information. Classification models with classifier-based wrapper feature selection were employed to estimate feature importance and identify diagnostically relevant acoustic characteristics. Among the evaluated classifiers, the extreme minimal learning machine emerged as the most computationally efficient method, providing competitive classification accuracy with substantially lower training time in repeated leave-one-subject-out validation. The results demonstrated that the proposed framework is computationally efficient, interpretable, and well-suited as a supportive tool for speech-based dementia assessment.
\end{abstract}


\maketitle

\section{Introduction}
\label{sec:Intro}

Dementia is a progressive clinical syndrome characterized by cognitive and functional decline that interferes with daily living \cite{pasquier1999early, world2017global}. It results from underlying diseases, most commonly Alzheimer's disease, which accounts for 60 to 70\% of cases, and its risk is associated with advanced age, genetic predisposition, and chronic conditions \cite{world2017global}. The disease progresses gradually from subtle impairments in memory, learning, and attention to severe deficits in language, executive function, and daily functioning \cite{world2017global,wang2019cognitive}. More than 55 million people are currently affected worldwide, and the number of cases is projected to nearly triple by 2050, emphasizing the importance of early and efficient recognition of cognitive decline \cite{world2017global}.

Standardized cognitive screening tools such as the Mini-Mental State Examination (MMSE) \cite{folstein1975mini}, the Montreal Cognitive Assessment (MoCA) \cite{nasreddine2005montreal}, and the Consortium to Establish a Registry for Alzheimer's Disease (CERAD) battery \cite{welsh1994consortium} are widely used in clinical practice. These assessments evaluate multiple cognitive domains, including memory, language, attention, and visuospatial abilities. Mobile cognitive screening applications have also been developed and shown to correlate with established tests \cite{zorluoglu2015mobile}.

Neuroimaging techniques such as MRI, PET, and CT, together with biomarker analyses including cerebrospinal fluid and blood tests, and clinical anamnesis, are utilized in dementia diagnostics \cite{jack2018nia}. In Alzheimer's disease, structural changes are particularly evident in memory-related regions such as the hippocampus and medial temporal lobe \cite{hanyu2000magnetization}. Machine learning, especially deep learning-based convolutional neural networks, has been applied to neuroimaging-based diagnosis \cite{ebrahimighahnavieh2020deep, mirzaei2022machine}. However, these approaches are often constrained by high acquisition costs, invasiveness, limited standardization, and restricted data availability \cite{young2020imaging}.

The analysis of naturally spoken language provides a cost-effective and non-invasive alternative for identifying dementia. In Alzheimer's disease and related disorders, early impairments may manifest as difficulties in semantic memory, word retrieval, and sentence construction \cite{reilly2011anomia,ivanova,fraser2015linguistic}. Advances in speech analysis have enabled the development of machine learning models that utilize acoustic and linguistic features to detect cognitive impairment \cite{de2020artificial}. Consequently, there is a growing need for interpretable and computationally efficient speech-based approaches that support accurate classification while enabling systematic analysis of the underlying acoustic characteristics.

Despite this growing interest, relatively few publicly available speech databases are available for the detection of cognitive decline. Although analyses are often comprehensive, experiments have typically been conducted with relatively small sample sizes, which may limit statistical power and reduce the generalizability of findings to larger populations \cite{de2020artificial}. Limited sample sizes are also a broader challenge in supervised classification settings, particularly when developing data-driven or deep learning-based models \cite{zhen2023}. Another common methodological concern is the imbalance between healthy control participants and individuals with dementia, as well as differences in age and gender distributions across groups. Such imbalances can bias classification models and distort performance estimates if evaluation metrics do not adequately account for class distribution. In addition, some studies have included multiple speech recordings from the same individuals in both training and testing sets, increasing the risk that models capture speaker-specific characteristics rather than markers of cognitive impairment. Patients have also been stratified according to dementia severity in some studies, making direct comparisons of results more difficult. Furthermore, many studies do not provide open access to their datasets \cite{de2020artificial}.

Taken together, these limitations indicate that systematic evaluation of compact and interpretable acoustic feature subsets under strictly controlled and subject-independent validation protocols remains limited. The purpose of this study is twofold: (i) to evaluate the potential of machine learning models to predict cognitive decline from human speech, and (ii) to identify the most important acoustic characteristics contributing to classification performance. Although deep learning techniques are increasingly used in this research area \cite{alsuhaibani2025review}, they were not included in the present study because there is currently no widely accepted method for directly linking their performance to specific acoustic characteristics \cite{ortiz2023deep,csahin2025unlocking}. In contrast, the interpretable classifiers used in this work enable an explicit wrapper-based approach for assessing feature importance in a systematic manner \cite{linja2023feature}. Furthermore, experiments reported by Kumar et al. \cite{kumar2022dementia} indicated that non-deep-learning-machine learning methods achieved higher dementia classification accuracy than a representative set of deep learning techniques.

This study utilizes the INTERSPEECH 2020 Alzheimer's Dementia Recognition through Spontaneous Speech (ADReSS) challenge dataset to classify cognitively healthy controls and individuals with Alzheimer's disease \cite{luz2021alzheimer}. Model performance is evaluated using leave-one-subject-out (LOSO) cross-validation and an independent test set following the original challenge protocol. In addition, an extended subset of recordings from the DementiaBank Pitt Corpus is used to assess the generalization of the proposed methods to a larger and more heterogeneous dataset.

Acoustic features are extracted using the open-source openSMILE toolkit. Classification is performed with classifier-based wrapper models including a linear support vector machine (L-SVM)~\cite{cortes1995support}, ridge logistic regression (Ridge) \cite{hastie2009elements}, and extreme minimal learning machine (EMLM) \cite{karkkainen2019extreme}, enabling efficient learning and estimation of feature weights. Each recording is represented by a single feature vector, reducing computational complexity compared to segment-level approaches. Informative acoustic features are identified by ranking model-assigned weights and evaluating classification performance across increasing feature subset sizes. In parallel, a permutation-based statistical cutoff is used to estimate a data-driven feature subset size.

The main contribution of this study is the systematic evaluation of recording-level acoustic feature representations combined with classifier-based wrapper feature selection for dementia classification. The present work emphasizes the identification and comparison of diagnostically relevant acoustic feature categories across multiple lightweight and interpretable classifiers.

\section{Background}
\label{sec:Background}

Acoustic features can be extracted directly from audio recordings without converting speech into text, enabling fast, resource-efficient, and cost-effective analysis \cite{luz2021alzheimer}. In dementia detection, acoustic approaches are particularly attractive because they are largely independent of language, educational background, and cultural factors.

Changes in cognitive function can manifest as alterations in speech production, prosody, and voice quality. In dementia, variations in fundamental frequency (F0) are commonly observed and may be reflected in increased voice jitter and amplitude shimmer \cite{meilan}. Reduced F0 variability can lead to monotonous speech with limited intonational modulation \cite{martinez}, while changes in voice intensity dynamics may indicate impairments in emotional expression and motor control \cite{scherer2003vocal}. Temporal aspects of speech are also affected: slower speech rates, irregular rhythm, and altered pause patterns have been associated with cognitive decline, reflecting difficulties in linguistic planning, working memory, and lexical retrieval \cite{pistono2016pauses,pappagari2020using}. In addition, spectral characteristics such as Mel-frequency cepstral coefficients (MFCCs) and formant frequencies (F1 and F2) capture changes in articulatory precision and voice quality, which may be degraded in dementia through increased hoarseness, breathiness, and reduced vocal stability \cite{fraser2015linguistic,meghanani2021exploration,parlak2023voice}.

To represent these speech characteristics, several standardized acoustic feature sets have been proposed. The Extended Geneva Minimalistic Acoustic Parameter Set (eGeMAPS) provides a compact collection of physiologically and perceptually motivated features, while larger sets such as the Computational Paralinguistics Challenge (ComParE) and COVAREP offer high-dimensional representations covering a broad range of prosodic, spectral, and voice-quality features \cite{eyben2015geneva,eyben2015realtime,degottex}. Although high-dimensional feature sets can capture rich information, they often require careful feature selection or dimensionality reduction to ensure interpretability and computational efficiency, particularly in clinical applications.

In parallel, deep learning-based approaches have been increasingly explored for acoustic representation learning. These include embedding-based methods such as x-vectors and VGGish, as well as end-to-end convolutional neural networks operating on raw or spectrogram-based speech representations \cite{snyder2018x,gemmeke2017audio,cummins2020comparison}. While such approaches can automatically learn task-specific representations and can benefit from pretrained speech representation models, they may still require substantial computational resources and careful adaptation, and their learned representations are often less directly interpretable than traditional acoustic feature-based methods \cite{baevski2020wav2vec,balagopalan2021comparing}.

A common alternative to deep learning is the aggregation of low-level descriptors (LLDs) extracted from short-time speech frames into global representations using statistical functionals~\cite{eyben2015realtime}. Related statistical embedding techniques, such as i-vectors, compress speech characteristics into low-dimensional representations using probabilistic modeling without relying on deep neural networks~\cite{dehak2010front}. More recently, distribution-based aggregation methods, including bag-of-audio-words (BoAW), fisher vectors (FV), and clustering-based representations, have been applied to summarize acoustic feature distributions for dementia classification, with recording-level BoAW representations combined with statistical feature selection showing competitive performance while maintaining interpretability and computational efficiency \cite{schmitt2017openxbow,syed2020automated,haider2019assessment,niemela2024classification}.

Multimodal approaches that combine acoustic and lexical information have also been proposed to improve diagnostic performance \cite{pappagari2020using,martinc2020tackling,rohanian2021multi}. However, such methods introduce additional linguistic dependencies and complexity.

In contrast to deep learning-based and multimodal approaches, the present study focuses on interpretable acoustic features aggregated at the recording level and combined with computationally efficient wrapper-based feature selection. This design choice enables accurate, transparent, and scalable dementia classification from spontaneous speech while reducing reliance on lexical content and potentially improving portability across languages.

\section{Methods}

\subsection{Pitt Corpus dataset}

The original Pitt Corpus, a core component of the DementiaBank project\footnote{\url{https://dementia.talkbank.org/access}}, was collected between~1983 and 1988 as part of Alzheimer's disease research \cite{becker1994natural}. A commonly analyzed subset of this corpus consists of 101 healthy control individuals and 181 patients with Alzheimer's disease. Participants' diagnoses were based on clinical assessments and long-term follow-up. Eligibility criteria included a minimum age of 44 years, at least seven years of education, the ability to read and write prior to the onset of dementia, the absence of central nervous system disorders or relevant medications, a Mini-Mental State Examination (MMSE) score of at least 10/30, and informed consent from both the participant and a caregiver.

Participants were asked to describe events depicted in the kitchen-scene picture from the Boston Diagnostic Aphasia Examination (BDAE) test battery \cite{fong2019factor}. A caregiver provided initial guidance at the beginning of the task and could ask follow-up questions or introduce additional topics during free-form conversation. All responses were recorded as audio files and later transcribed using the CHAT format \cite{macwhinney2014childes}, including annotations for pauses, repetitions, and errors.

While the original Pitt subset of the DementiaBank corpus is frequently analyzed in computational studies, the project has since released expanded and updated versions of the dataset. These later releases include additional participants as well as revised diagnostic information, which explains the discrepancies in reported sample sizes across studies.

Although the primary experimental evaluation in this study is conducted using the ADReSS 2020 challenge dataset\footnote{\url{https://dementia.talkbank.org/ADReSS-2020/}}, the extended Pitt Corpus\footnote{\url{https://dementia.talkbank.org/access/English/Pitt.html}} is analyzed as a complementary dataset to assess the robustness of the proposed methods on a larger and more heterogeneous sample. In this work, we use the available extended Pitt Corpus data, consisting of the original Pitt recordings together with subsequent Pitt Corpus additions and diagnostic updates.

In total, the extended Pitt corpus includes recordings from 104 cognitively healthy control participants and 208 participants with dementia. During preprocessing, a subset of recordings was excluded due to insufficient audio quality or because multiple recordings were available for the same participant. In the latter case, a single representative recording was retained, resulting in one recording per participant and a final dataset of 291 recordings.

The final control group consisted of 99 individuals, of whom 58 were female (mean age~=~65.3~$\pm$~8.2 years) and 41 were male (mean age = 65.6 $\pm$ 8.0 years). The dementia group consisted of 192 individuals, including 125 females (mean age = 72.4 $\pm$ 8.6 years) and 67 males (mean age~=~70.1~$\pm$~8.5 years).

\subsection{ADReSS 2020 dataset}
\label{sec:ADReSS}

The main goal of the ADReSS dataset was to provide a standardized and openly available speech database to enable the evaluation and comparison of different machine learning models for detecting cognitive decline \cite{luz2021alzheimer}. The dataset is a subset of the Pitt Corpus consisting of 156 individuals, containing one recording per individual. The participants were selected so that half were healthy control individuals and half dementia patients. Additionally, the dataset is divided into a training set of~108 individuals (54 healthy controls and 54 with Alzheimer's disease) and a test set of 48 individuals (24 healthy controls and 24 with Alzheimer's disease). Table \ref{Table1:ADReSS_Set} illustrates the distribution, taking into account the age and gender distribution of the participants.

The classification task distinguishes participants diagnosed with Alzheimer's disease/dementia from cognitively healthy controls. In the ADReSS dataset, the diagnostic group is referred to as \emph{Alzheimer's disease}, whereas in the extended Pitt Corpus, we use the term \emph{dementia} following the available corpus metadata.

\begin{table}[H]
\centering
\caption{ADReSS 2020 dataset (M = male, F = female, AD = Alzheimer's disease).} \label{Table1:ADReSS_Set}  
\begin{tabular}{lllllllll}
\hline
Age& \multicolumn{2}{c}{Train-AD} & \multicolumn{2}{c}{Test-AD} & \multicolumn{2}{c}{Train-Non-AD} & \multicolumn{2}{c}{Test-Non-AD} \\
 & M  & F & M & F & M & F & M & F \\
\hline
{[}50, 55) & 1 & 0 & 1 & 0 & 1 & 0 & 1 & 0 \\
{[}55, 60) & 5 & 4 & 2 & 2 & 5 & 4 & 2 & 2 \\
{[}60, 65) & 3 & 6 & 1 & 3 & 3 & 6 & 1 & 3 \\
{[}65, 70) & 6 & 10 & 3 & 4 & 6 & 10 & 3 & 4 \\
{[}70, 75) & 6 & 8 & 3 & 3 & 6 & 8 & 3 & 3 \\
{[}75, 80) & 3 & 2 & 1 & 1 & 3 & 2 & 1 & 1 \\
\hline
Total & 24 & 30 & 11 & 13 & 24 & 30 & 11 & 13 \\
\hline
\end{tabular}
\end{table}

\subsection{Data preprocessing}

The ADReSS dataset provides audio recordings segmented into active speech intervals. In the present study, however, acoustic features were extracted from entire recordings obtained from the DementiaBank database hosted by the TalkBank project. This choice enabled a compact recording-level representation, with one feature vector per participant, thereby reducing computational complexity compared with segment-level approaches. Because the analysis was performed at the level of entire recordings, pause, hesitation, and other paralinguistic information may also be reflected in the resulting acoustic feature representations.

The recordings contained substantial background noise and extraneous sounds, including caregiver speech, buzzer sounds, traffic noise, phone ringing, and sudden environmental noises. Initial noise reduction was performed using the adaptive noise reduction filter in \platfont{Adobe Audition (version 24.4, 64-bit)}, which did not require manual parameter tuning. To improve consistency in manual preprocessing, the same criteria were applied throughout: clearly non-participant and non-task-related sounds were removed, whereas participant speech, silent pauses, and paralinguistic vocalizations were retained.

The recordings were originally sampled at 44 kHz and subsequently downsampled to 16 kHz, which allows reliable signal representation within the 0--8 kHz frequency range. As most speech-related frequency components fall below 8 kHz, a 16-kHz sampling rate is sufficient for speech analysis. All recordings were normalized according to the EBU R 128 loudness standard \cite{ebu2011loudness} to reduce variability arising from recording conditions, such as microphone placement. The processed audio files were stored in WAV format.

In the extended Pitt Corpus, multiple recordings were available for several participants. One recording per participant was selected based on the highest speech intelligibility in noise (SIG) score to ensure consistent audio quality. The SIG score provides an objective estimate of speech signal quality in noisy conditions, following the ITU-T Recommendation P.835 \cite{itutp835}.

For the ADReSS dataset, the average recording duration did not differ significantly between control participants (mean = 51.9 s, SD = 24.9 s) and participants with Alzheimer's disease (mean = 58.7 s, SD = 33.0 s), as indicated by a two-sample t-test ($p > 0.05$). In the extended Pitt Corpus, the mean recording duration was 47.4 s (SD = 21.2 s) for control participants and 56.1 s (SD = 29.9 s) for participants with dementia. This difference was statistically significant at the 0.05 level ($p < 0.05$), but not at the 0.01 level. Recording duration was not included as an input feature in the classification models.

\subsection{Feature extraction}

Acoustic feature extraction was performed using the default settings of the \platfont{openSMILE (version~3.0)} library\footnote{\url{https://www.audeering.com/research/opensmile/}}. Three widely used acoustic feature sets were extracted: 88 eGeMAPS features,~988 EmoBase features, and 6373 ComParE features \cite{eyben}. These sets were combined into a single acoustic feature space comprising 7449 features. From this combined set, duplicate features arising from the overlap between the EmoBase and ComParE feature sets were first removed (371 features). Subsequently, features exhibiting very low variability (fewer than 10 unique values across the dataset) were removed (155 in total) to avoid including near-constant or highly quantized variables. The resulting feature space contained 6923 acoustic features.

The retained features primarily consist of low-level spectral descriptors, their first-order temporal derivatives (delta coefficients), and statistical functionals computed over these descriptors \cite{eyben2015realtime}. In addition to standard statistical measures such as mean, standard deviation, and minimum and maximum values, the feature set includes higher-order distributional statistics (e.g., percentiles, interquartile ranges, skewness, and kurtosis) as well as commonly used spectral and temporal descriptors. These include measures related to pitch dynamics, regression-based trend estimates, spectral variability and entropy, spectral centroid, and amplitude-related characteristics, which are widely employed in paralinguistic speech analysis.

\subsection{Classification and feature selection}

Next, we will present the classification methods used in this study, as well as the wrapper-type feature selection methods based on them. The three classifiers were selected to represent complementary lightweight and interpretable modeling strategies. L-SVM and Ridge provide linear decision functions with explicit coefficient-based feature weights, whereas EMLM provides a distance-based formulation with sensitivity-based feature importance estimates. Using these complementary models enables comparison of whether similar acoustic feature categories are emphasized across different classifier formulations while maintaining low computational cost.

\subsubsection{Linear support vector machine}

A linear support vector machine seeks a separating hyperplane that maximizes the margin between classes while allowing for misclassifications through a soft-margin formulation \cite{cortes1995support}. The model is defined by the linear decision function:
\[
f(\mathbf{x}_i) = \mathbf{w}^{\top}\mathbf{x}_i + b,
\]
where $\{\mathbf{x}_i\}_{i=1}^N$ denotes the observation vectors with $\mathbf{x}_i \in \mathbb{R}^n$, where $n$ represents the number of features, $N$ is the total number of observations, $\mathbf{w} \in \mathbb{R}^n$ is the weight vector, $b$ is the bias term, and $y_i \in \{-1,1\}$ denotes the class label. The model parameters are obtained by minimizing the objective function:
\[
\mathcal{J}_{L-SVM}(\mathbf{w}, b) = \frac{1}{2} \|\mathbf{w}\|^2 + C \sum_{i=1}^{N} \max(0, 1 - y_i f(\mathbf{x}_i)),
\]
where $C$ controls the trade-off between margin maximization and classification error. In the context of this study, the L-SVM is employed as a wrapper model that provides explicit feature weight estimates. Feature weights are given by the learned components of the weight vector $\mathbf{w}$, whose magnitudes reflect the contribution of each feature to the decision boundary and can therefore be used to rank acoustic features.

\subsubsection{Logistic ridge regression}

Logistic ridge regression (Ridge) extends standard logistic regression by incorporating L2 regularization to control model complexity and reduce overfitting \cite{hastie2009elements}. The probability of class membership is modeled using the logistic function:
\[
\hat{p}_i = \frac{1}{1 + e^{-f(\mathbf{x}_i)}},
\]
where $f(\mathbf{x}_i) = \mathbf{w}^{\top}\mathbf{x}_i + b$ denotes the linear predictor of the logistic regression model. Model parameters are learned by minimizing the regularized negative log-likelihood:
\[
\mathcal{J}_{Ridge}(\mathbf{w}, b) = - \sum_{i=1}^{N} \left[y_i \log(\hat{p}_i) + (1 - y_i)\log(1 - \hat{p}_i) \right] + \frac{\lambda}{2}\|\mathbf{w}\|^2,
\]
where the regularization parameter $\lambda \geq 0$ controls the magnitude of the coefficients in the weight vector and $y_i \in \{0,1\}$ denotes the class label. Optimization is performed using gradient-based numerical methods. Feature weights correspond to the learned coefficients $\mathbf{w}$, which are used to rank acoustic features in the wrapper-based feature selection procedure.

\subsubsection{Extreme minimal learning machine}

The extreme minimal learning machine is a distance-based supervised machine learning model \cite{karkkainen2019extreme}. It combines ridge regression with the distance-based representation of the minimal learning machine (MLM), which is based on distance computations \cite{karkkainen2019extreme,hamalainen2020minimal}. The corresponding model output function can be written as 
\[
f_{\mathrm{EMLM}}(\mathbf{x}_i) = \mathbf{W}\mathbf{H}(\mathbf{x}_i),
\]
where $\mathbf{W} \in \mathbb{R}^{k\times m}$ denotes the matrix of weight coefficients for distance-based classification, $k$ is the number of classes, $m$ is the number of reference points, and $\mathbf{H}(\mathbf{x}_i)$ is the vector of distances between the reference points $\{\mathbf{r}_j\}_{j=1}^m$ and the observation $\mathbf{x}_i$. For a training dataset consisting of $N$ observations, the distances between each observation and the $m$ reference points are collected into the matrix $\mathbf{H}$, where $\mathbf{H} \in \mathbb{R}^{m\times N}$ and $(\mathbf{H})_{ji} = \lVert \mathbf{r}_j - \mathbf{x}_i \rVert_2$. 

The weight matrix $\mathbf{W}$ is obtained by solving a ridge-type regularized least-squares problem, in which the regularization term stabilizes the solution and ensures unique solvability \cite{karkkainen2019extreme,karkkainen2019model}. This leads to the linear matrix equation:
\[
\mathbf{W}\left(\mathbf{H}\mathbf{H}^T + \frac{\lambda_{\mathrm{EMLM}} N}{m}\mathbf{I}\right) = \mathbf{Y}\mathbf{H}^T.
\]
Following \cite{karkkainen2019extreme}, we fix $\lambda_{\mathrm{EMLM}} = \sqrt{\epsilon_{\mathrm{mach}}}$, where $\epsilon_{\mathrm{mach}}$ denotes machine epsilon, because $\lambda_{\mathrm{EMLM}}$ is used to ensure unique solvability rather than to control model complexity. 

For the full EMLM model considered here, all $N$ observations themselves are used as reference points, i.e., $m = N$. The columns of the matrix $\mathbf{Y} \in \mathbb{R}^{k\times N}$ contain the 1-of-$k$ encoded class labels. For a new observation $\mathbf{x}^*$, the predicted output of the trained model is given by $\mathbf{y}^* = \mathbf{W}\mathbf{H}^*,$ where $\mathbf{H}^* = [H_1^*,\dots,H_m^*]^T$ is the vector of distances between the reference points and the observation $\mathbf{x}^*$, and $H_j^* = \lVert \mathbf{r}_j - \mathbf{x}^* \rVert_2,\quad j=1,\dots,m.$ The output values can be interpreted as class-wise scores for the observation. In the binary case, the predicted class corresponds to the larger output value.

For the input feature vectors in the EMLM model's training dataset, the average absolute sensitivity values can be computed from the gradient of the model output function and used as feature importance weights to rank the acoustic features. The sensitivity values quantify the effect of the input features on the model output and are obtained from the gradient of the output function with respect to the input features:
\[
\mathcal{FI} = \frac{1}{N} \sum_{i=1}^N\left|\diffp{f_{\mathrm{EMLM}}}{\mathbf{x}_i}\right|.
\]
The gradient of the output can be expressed in a well-defined form \cite{karkkainen2005computation} as follows:
\[
\diffp{f_{\mathrm{EMLM}}}{\mathbf{x}_i} = \mathbf{W}\mathbf{D}^T, \quad \mathbf{D}(\mathbf{x}_i) = [\mathbf{d}_1,\mathbf{d}_2,\dots,\mathbf{d}_m], \quad \mathbf{D}(\mathbf{x}_i) \in \mathbb{R}^{n \times m},
\]
\[
\mathbf{d}_j = \frac{\mathbf{r}_j-\mathbf{x}_i} {\max\left(\epsilon_{\mathrm{dist}},\lVert \mathbf{r}_j-\mathbf{x}_i\rVert_2\right)}, \quad \mathbf{d}_j \in \mathbb{R}^{n}, \quad j=1,\dots,m.
\]
Here, $\epsilon_{\mathrm{dist}}$ is set to $\sqrt{\epsilon_{\mathrm{mach}}}$ to avoid division by zero or numerical instability when the distance between a reference point and an observation is very small.

\subsection{Experimental evaluation and validation setup}

Classification performance was evaluated using two complementary settings, following the protocol described in \cite{luz2021alzheimer}: LOSO cross-validation and a separate test dataset. LOSO cross-validation was conducted on both the ADReSS dataset and the extended Pitt Corpus, while the separate test dataset comprised approximately one-third of the ADReSS dataset (48 of 156 individuals) and was used to assess generalization to unseen data. In all experiments, min-max normalization to the range $[0, 1]$ was applied to the training data, and the same normalization coefficients estimated from the training data were subsequently applied to the corresponding test data.

Accuracy was used as the primary performance metric because the ADReSS training and test sets are class-balanced. In addition, class-specific F1-scores, precision, and recall were reported to account for class imbalance in the extended Pitt Corpus and to evaluate whether the models differed in their sensitivity to AD/dementia and Non-AD/control groups.

As LOSO cross-validation requires training the classification model once for each subject, its overall computational complexity scales linearly with the number of subjects and can be expressed as
$\mathcal{O}\bigl(N \cdot C_{\mathrm{train}}\bigr),$
where $N$ denotes the number of subjects and $C_{\mathrm{train}}$ denotes the computational cost of training a single model.

All experiments were conducted using a consistent computational environment to ensure comparability across methods. Specifically, experiments were run on a Linux workstation operating \platfont{Rocky Linux 9.7} and equipped with an \platfont{Intel Xeon Platinum 8160 CPU (2.10 GHz, 48 physical cores)}. The implementations relied on \platfont{MATLAB R2023b (Update 6)} and \platfont{Python 3.9.25}, depending on the classifier, and the glmnet MATLAB binaries were compiled for the Linux platform using default compilation settings.

Three classification models were evaluated. For the EMLM model, both feature weighting and classification were conducted using all observations as reference points ($m=N$). The implementation was based on publicly available code\footnote{\url{https://gitlab.jyu.fi/hnpai-public/extreme-minimal-learning-machine}}.

The L-SVM model was implemented using the \platfont{scikit-learn (v.1.6.1)} library with the LIBLINEAR backend\footnote{\url{https://scikit-learn.org/stable/api/sklearn.svm.html}}. 
During training in feature weighting, the default regularization parameter ($C=1.0$) was used. For model training and classification, a smaller regularization value ($C=0.01$) was applied.

The Ridge model was implemented as L2-regularized logistic regression using the \platfont{Glmnet for Matlab (2013)} library\footnote{\url{https://hastie.su.domains/glmnet_matlab/}} \cite{hastie2014glmnet}. 
During training in feature weighting, the default regularization parameter ($\lambda=1.0$) was used, whereas model training and classification employed a lower regularization strength ($\lambda=0.1$).

Preliminary experiments were also conducted using a random forest classifier \cite{breiman2001random}. However, due to the small sample size and high feature dimensionality, this approach exhibited substantial overfitting even under strong regularization settings ($\textit{MinLeafSize}=10$, $\textit{MaxNumSplits}=5$) and was therefore excluded from the final experiments.

The primary objective of the experiments was to evaluate classification performance using wrapper-based feature ranking derived from the three models. Feature weights were estimated exclusively using the ADReSS training data following Algorithm \ref{Algorithm:Alg1} and were subsequently applied in all classification experiments. Feature ranking was performed once using the ADReSS training set and was not nested within the LOSO folds. For each classifier, features were ranked according to their model-specific weights, and classification was first performed using the single highest-weighted feature. The number of features was then increased incrementally by adding one feature at a time, with classification repeated at each step.

As a secondary objective, the number of features required for stable classification performance was estimated separately for each classifier using a one-sided Wilcoxon signed-rank test \cite{gibbons2021nonparametric}. The resulting ranked feature lists and statistically determined cutoffs were subsequently used in the classification experiments described in the following section (Section \ref{sec:permutationTesting}).

\begin{algorithm}[H]
	\caption{
		 Repeated feature weight estimation and ranking.}
	\label{Algorithm:Alg1}
	\begin{algorithmic}[1]	
		\FOR{$i=1$ to $100$}
		\STATE Randomly split the data into 80\% training and perform 5-fold cross-validation.
		\FOR{each fold}
		\STATE Normalize training data (min-max), train the model, and compute feature weights using true class labels.
		\STATE Permute class labels and recompute feature weights.
		\ENDFOR
		\STATE Compute median feature weights across folds for true and permuted labels.
		\STATE Take absolute values of the median weights for true and permuted labels.
		\ENDFOR
		
		\STATE Average the absolute median weights (true labels) across iterations.
		\STATE Rank features according to the averaged absolute median weights (true labels).
		
		\STATE \textbf{Output:} Ranked feature list and distributions of absolute median weights for true and permuted labels.
	\end{algorithmic}
\end{algorithm}

\subsection{Feature ranking and permutation-based subset selection}
\label{sec:permutationTesting}

In this study, the feature ranking procedure is used as an application of classifier-based wrapper feature selection, rather than as a new general-purpose feature selection method. The statistical significance and importance ranking of the feature weights obtained from the classifiers were estimated using a repeated random subsampling and cross-validation framework, as summarized in Algorithm \ref{Algorithm:Alg1}. Feature weight estimation was conducted using the training portion of the ADReSS dataset, comprising feature vectors from 108 individuals and 6923 features. The data were randomly split into 80\% training data and evaluated using 5-fold cross-validation, repeated over 100 iterations. Within each fold, features were normalized using min-max scaling to the range $[0,1]$, based on training-fold statistics. The model was trained using true class labels to obtain feature weights, after which class labels were permuted and the model was retrained to obtain permuted weights. Median feature weights across folds were computed for both true and permuted labels, yielding 100 median estimates per feature. Features were ranked based on the average of the absolute values of the fold-wise median true-label weights.

To determine a cutoff for feature selection, the Wilcoxon signed-rank test was applied to the 100 paired absolute median weight estimates for each feature, comparing true-label and permuted-label weights (see Algorithm \ref{Algorithm:Alg2}). The null hypothesis assumed a non-positive median difference, and a one-sided (right-tailed) Wilcoxon signed-rank test was applied at $\alpha=0.10$. The significance level $\alpha = 0.10$ was selected to allow greater sensitivity in this exploratory analysis. In the present feature selection setting, this choice was intended to reduce the risk of prematurely discarding potentially informative acoustic features before evaluating their contribution to classification performance. The ranked list was traversed from the highest-weighted feature downward. The first feature for which the null hypothesis was not rejected was defined as the cutoff, and all higher-ranked features were selected. Similar data-driven feature count selection has been reported in previous studies \cite{jauhiainen2021new,saarela2021comparison}.

\begin{algorithm}[H]
\caption{
	Permutation-based feature subset selection.}
\label{Algorithm:Alg2}
\begin{algorithmic}[1]
\FOR{features in ranked order}
    \STATE Compute paired differences 
    $d_i = |w^{\text{true}}_i| - |w^{\text{perm}}_i|$, $i = 1,\dots,100$,
    \STATE \hspace{1em} where $w^{\text{true}}_i$ and $w^{\text{perm}}_i$ denote the absolute fold-wise median weights
    \STATE \hspace{1em} obtained in iteration $i$.
    \STATE Apply a one-sided Wilcoxon signed-rank test to $\{d_i\}_{i=1}^{100}$ 
    \STATE \hspace{1em} testing $H_0\!:\mathrm{median}(d_i)\le 0$ at $\alpha=0.10$.
    \IF{$H_0$ is not rejected}
        \STATE Stop and define the current feature as the cutoff.
    \ENDIF
\ENDFOR

\STATE \textbf{Output:} Features ranked above the cutoff.

\end{algorithmic}
\end{algorithm}
	\vspace{-8mm}
	
\section{Results}	\vspace{-2mm}
\subsection{Classification results}
\label{sec:classificationResults}
\begin{figure}[H]
	\centering	
	\begin{subfigure}{0.45\textwidth}
		\centering
		\includegraphics[width=\textwidth]{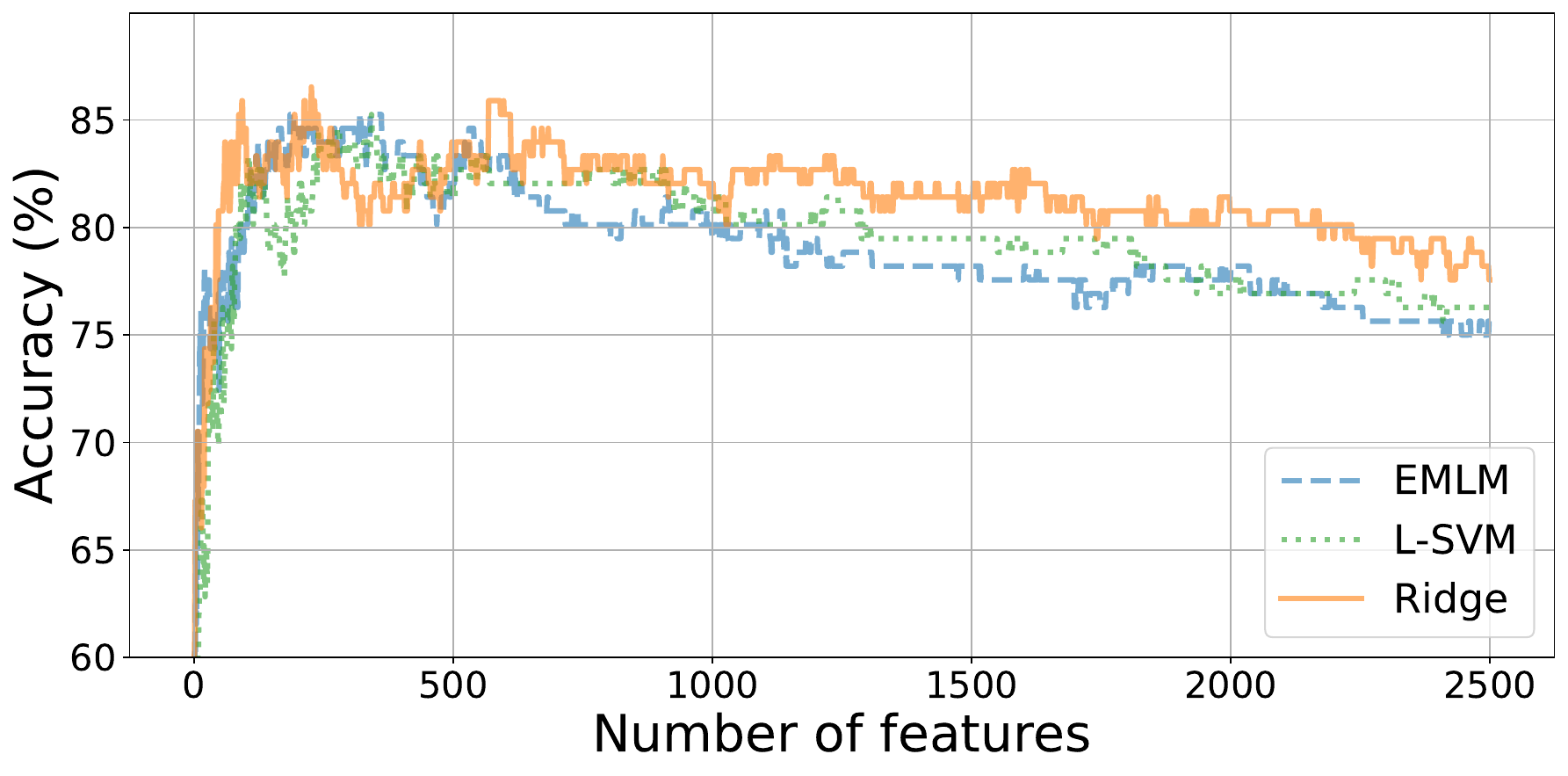}
	\caption{LOSO cross-validation (ADReSS dataset).}
	\end{subfigure}	
	\begin{subfigure}{0.45\textwidth}
		\centering
		\includegraphics[width=\textwidth]{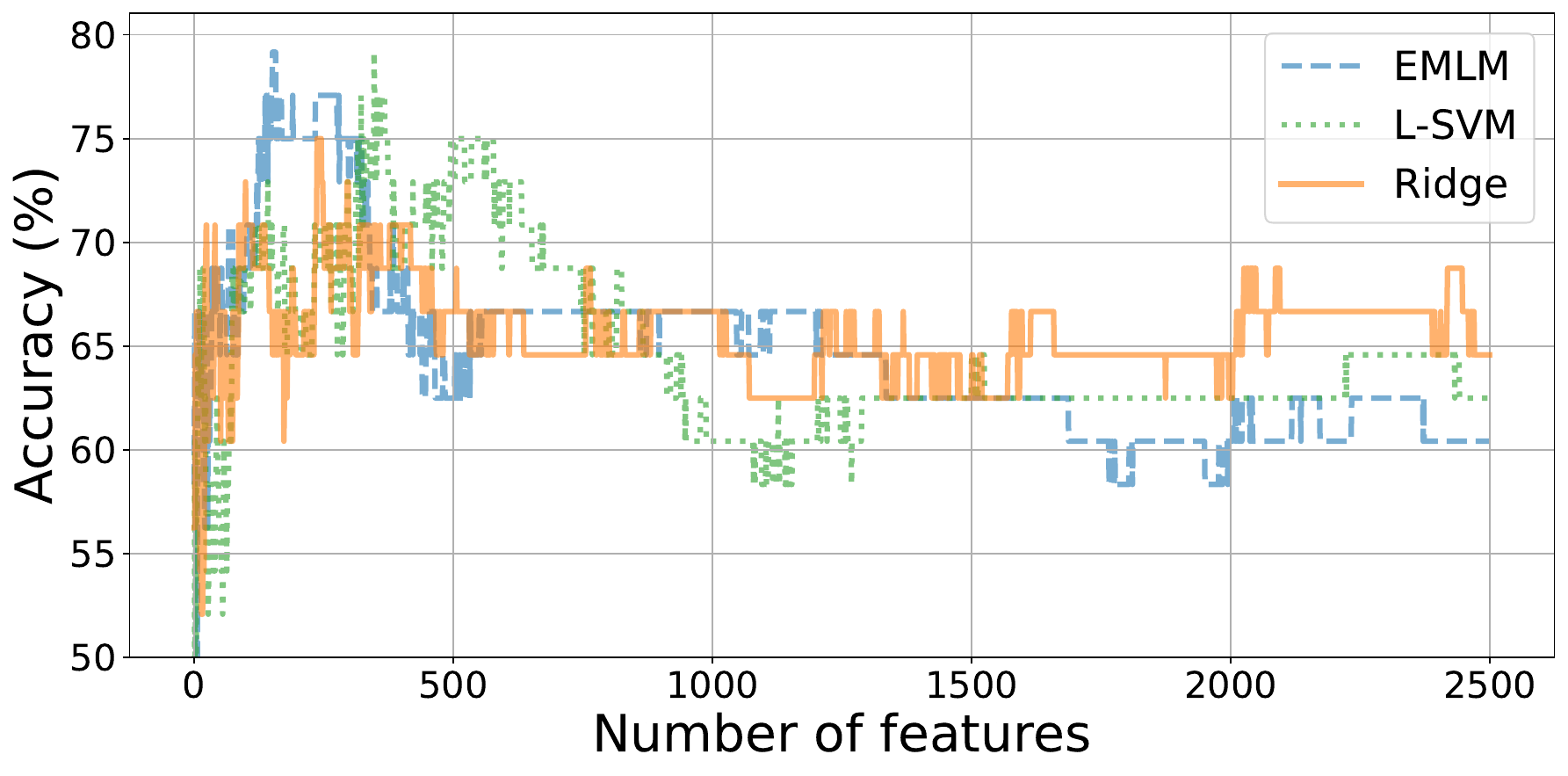}
	\caption{Independent test set.}
	\end{subfigure}	
	\begin{subfigure}{0.45\textwidth}
		\centering
		\includegraphics[width=\textwidth]{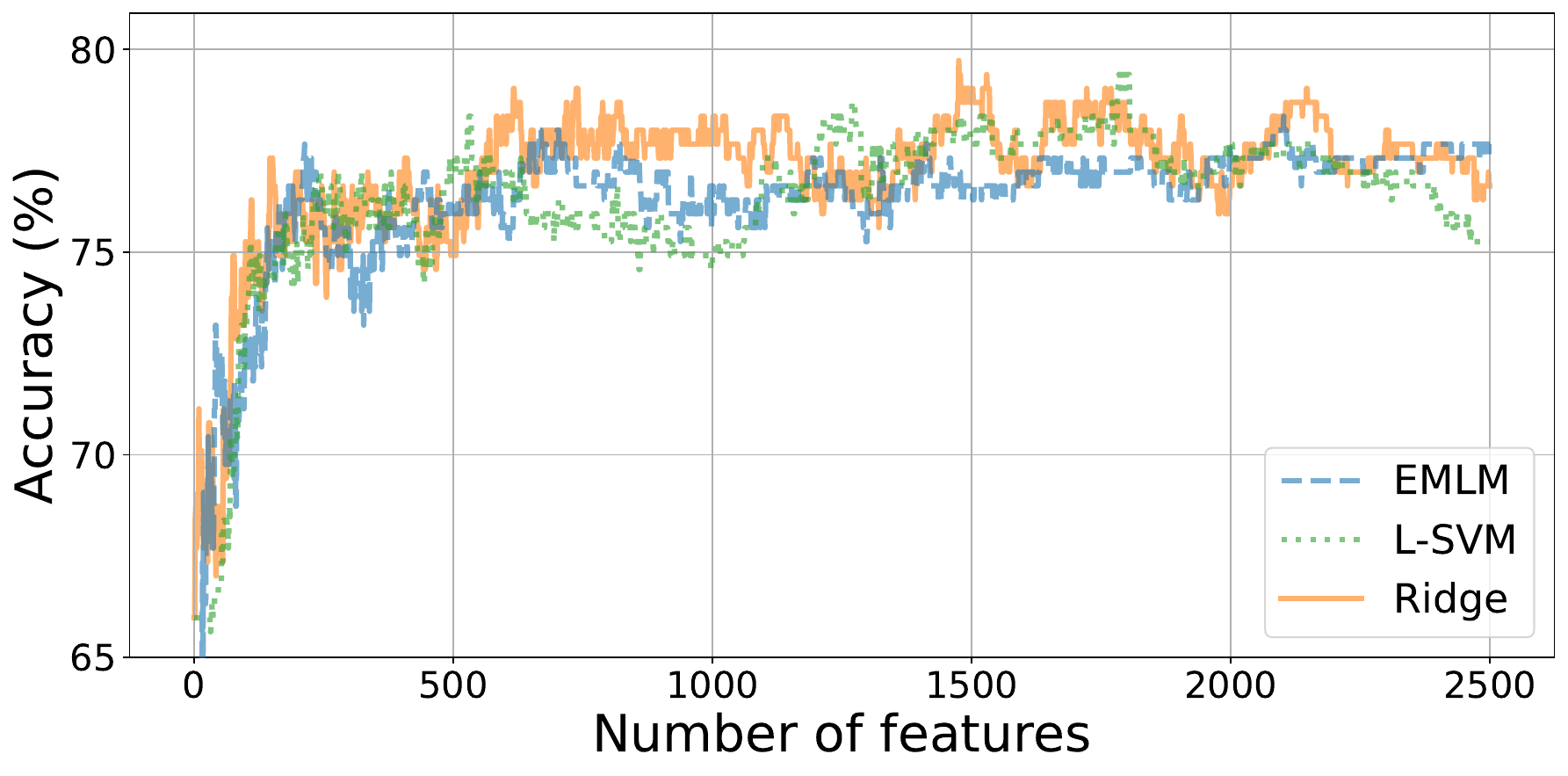} 
	\caption{LOSO cross-validation (extended Pitt Corpus).}
	\end{subfigure}	
	\vspace{-2mm}	\caption{Classification accuracy across datasets as a function of feature subset size, using up to the 2500 highest-weighted wrapper-selected features.}
	\label{Fig1:Accuracies_All}
\end{figure}

Figure \ref{Fig1:Accuracies_All} presents classification accuracies for the EMLM, L-SVM, and Ridge models across different evaluation settings, using wrapper-based feature weighting embedded within the classifiers and evaluated across feature subset sizes up to the 2500 highest-ranked features. Table \ref{Table2:ClassificationResults} summarizes classification performance in LOSO cross-validation on ADReSS, on the separate test dataset, and in LOSO cross-validation on the extended Pitt Corpus. Because the relationship between feature subset size and classification performance is non-monotonic, the best observed classification accuracy was identified by evaluating performance across all feature subset sizes, incrementally increasing the number of features from one up to 2500.
	\vspace{-5mm}

\begin{table}[H]
	\centering
	\caption{Best classification performance across different datasets.}
	\label{Table2:ClassificationResults}
	\begin{tabular}{lllll}
		\hline
		\multicolumn{5}{c}{\textbf{LOSO cross-validation (ADReSS)}} \\
		\hline
		Method & Accuracy & F$_1$ (Non-AD) & F$_1$ (AD) & Selected features \\
		\hline
		EMLM  & 85.3\% & 85.4\% & 85.2\% & 186 \\
		L-SVM & 85.3\% & 85.4\% & 85.2\% & 343 \\
		Ridge & 86.5\% & 86.5\% & 86.6\% & 227 \\
		\hline
		\multicolumn{5}{c}{\textbf{Separate test dataset}} \\
		\hline
		Method & Accuracy & F$_1$ (Non-AD) & F$_1$ (AD) & Selected features \\
		\hline
		EMLM  & 79.2\% & 79.2\% & 79.2\% & 152 \\
		L-SVM & 79.2\% & 78.3\% & 80.0\% & 348 \\
		Ridge & 75.0\% & 75.0\% & 75.0\% & 238 \\
		\hline
		\multicolumn{5}{c}{\textbf{LOSO cross-validation (extended Pitt Corpus)}} \\
		\hline
		Method & Accuracy & F$_1$ (Non-AD) & F$_1$ (AD) & Selected features \\
		\hline
		EMLM  & 78.4\% & 64.8\% & 84.4\% & 2103 \\
		L-SVM & 79.4\% & 67.4\% & 84.9\% & 1785 \\
		Ridge & 79.7\% & 69.1\% & 84.9\% & 1476 \\
		\hline
	\end{tabular}
\end{table}

In the ADReSS dataset, EMLM and L-SVM achieved comparable LOSO classification accuracy (85.3\%), with EMLM requiring a smaller feature subset, while Ridge yielded the highest accuracy (86.5\%). On the separate test dataset, EMLM and L-SVM reached the highest accuracy (79.2\%), whereas Ridge showed lower performance (75.0\%). On the extended Pitt Corpus, all models attained similar LOSO accuracy (78.4--79.7\%), with Ridge achieving the highest accuracy while relying on the smallest feature subset. Overall, classification accuracy was higher in LOSO cross-validation on ADReSS than on the separate test dataset or the extended Pitt Corpus.

\begin{figure}[H]
  \centering
  \begin{subfigure}{0.25\textwidth}
    \includegraphics[width=\textwidth]{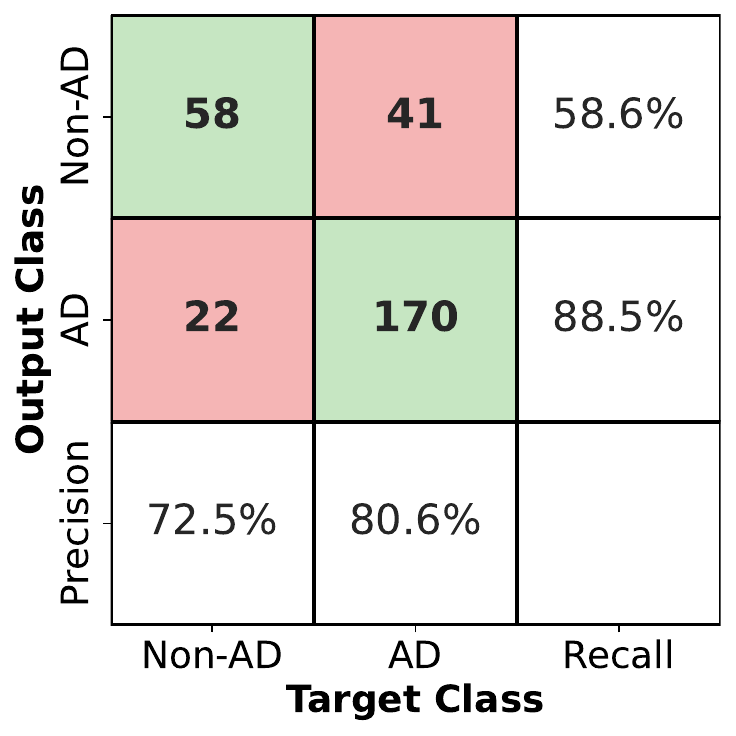} 
    \subcaption{EMLM}  
  \end{subfigure}
  ~
    \begin{subfigure}{0.25\textwidth}
    \includegraphics[width=\textwidth]{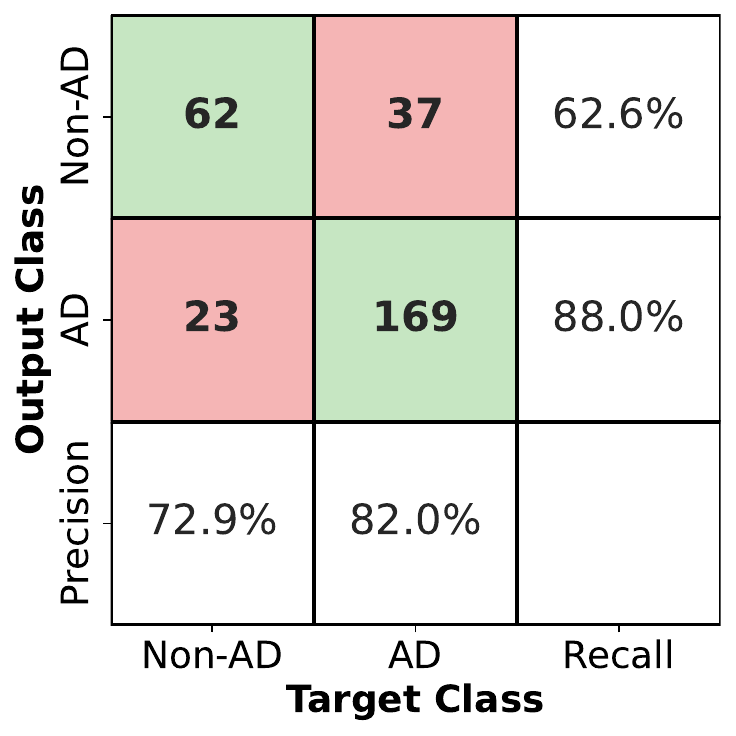}
    \subcaption{L-SVM} 
  \end{subfigure}
  ~
    \begin{subfigure}{0.25\textwidth}
    \includegraphics[width=\textwidth]{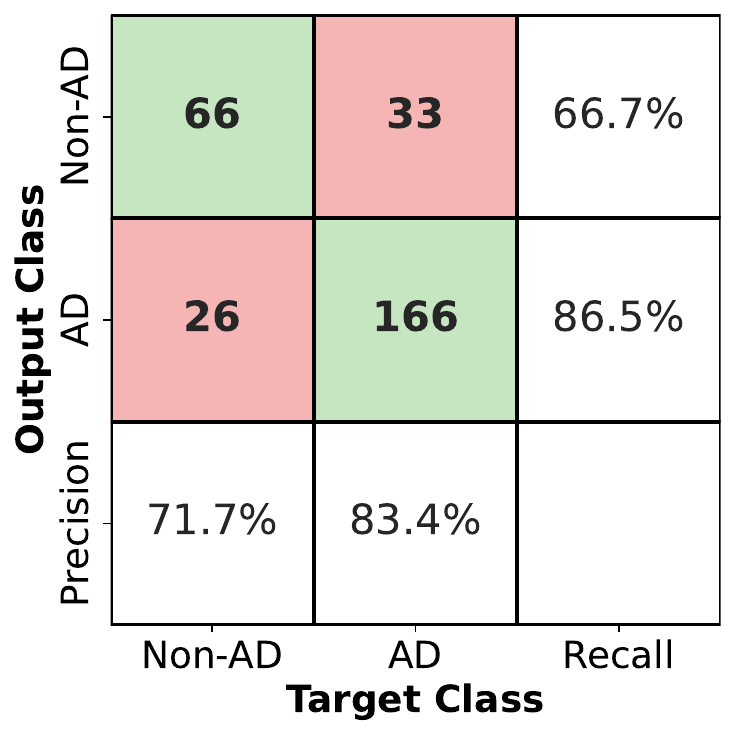} 
    \subcaption{Ridge}    
  \end{subfigure}
        \caption{Confusion matrices of the best-performing models, along with class-specific precision and recall.}
  \label{Fig2:PittCorpus_Confusion}
\end{figure}

Figure \ref{Fig2:PittCorpus_Confusion} shows the confusion matrices for the three classifiers on the extended Pitt Corpus dataset. Across models, recall for the AD class was consistently high (86.5--88.5\%), whereas recall for the Non-AD class was lower (58.6--66.7\%), indicating a higher rate of false positive classifications among Non-AD subjects. Precision followed a similar pattern, with higher values for the AD class (80.6--83.4\%) than for the Non-AD class (71.7--72.9\%). Overall, all classifiers exhibited higher sensitivity to AD than specificity to Non-AD, with Ridge showing the most balanced precision--recall trade-off across classes.

	\vspace{-3mm}

\subsection{Statistical estimation of feature subset size}
\label{sec:ResultsStatisticalEstimation}

The number of features was estimated separately for each classifier using a one-sided Wilcoxon signed-rank test (Table \ref{Table3:WilcoxonResults}). In LOSO cross-validation on the ADReSS dataset, all three classifiers achieved high classification accuracy (80.1--84.0\%) when using statistically selected feature subsets. When evaluated on the separate test dataset, classification performance decreased for all models, with accuracies of 75.0\%, 58.3\%, and 64.6\% for EMLM, L-SVM, and Ridge, respectively. On the extended Pitt Corpus, classification performance was lower than on ADReSS but remained relatively stable across models, with LOSO accuracy ranging from 75.3\% to 77.7\%. Overall, Wilcoxon-based feature selection yielded classification performance that was broadly comparable to that obtained through exhaustive empirical evaluation of feature subset sizes, although some reduction in accuracy was observed for certain classifiers, particularly on the independent test dataset.

\begin{table}[H]
\centering
\caption{Classification performance using features selected with the one-sided Wilcoxon signed-rank test ($\alpha = 0.10$) across different datasets.}
\label{Table3:WilcoxonResults}
\begin{tabular}{lllll}
\hline
\multicolumn{5}{c}{\textbf{LOSO cross-validation (ADReSS)}} \\
\hline
Method & Accuracy & F$_1$ (Non-AD) & F$_1$ (AD) & Selected features \\
\hline
EMLM  & 84.0\% & 84.1\% & 83.9\% & 279 \\
L-SVM & 80.1\% & 80.5\% & 79.7\% & 1120 \\
Ridge & 84.0\% & 84.5\% & 83.4\% & 642 \\
\hline
\multicolumn{5}{c}{\textbf{Separate test dataset}} \\
\hline
Method & Accuracy & F$_1$ (Non-AD) & F$_1$ (AD) & Selected features \\
\hline
EMLM & 75.0\% & 75.0\% & 75.0\% & 279 \\ 
L-SVM & 58.3\% & 56.5\% & 60.0\% & 1120 \\ 
Ridge & 64.6\% & 62.2\% & 66.7\% & 642 \\ 
\hline
\multicolumn{5}{c}{\textbf{LOSO cross-validation (extended Pitt Corpus)}} \\
\hline
Method & Accuracy & F$_1$ (Non-AD) & F$_1$ (AD) & Selected features \\
\hline
EMLM  & 75.3\% & 60.0\% & 82.1\% & 279 \\
L-SVM & 77.3\% & 62.5\% & 83.7\% & 1120 \\
Ridge & 77.7\% & 66.7\% & 83.2\% & 642 \\
\hline
\end{tabular}
\end{table}

\subsection{Feature category enrichment across classifiers}

Table \ref{Table4:FeatureCounts} summarizes the distribution of acoustic feature categories in the full feature set ($N = 6923$) and in the 500 most important features consistently identified across all three classifiers, based on average feature ranks. Enrichment was defined as the ratio between the relative frequency of a feature category in the selected feature subset and its relative frequency in the full feature set, reflecting the relative over-representation ($Enrichment>1$) or under-representation ($Enrichment<1$) of each category. Appendix 1 lists the top 100 features ranked by their average ranking position across classifiers, while Appendix 2 reports classifier-specific counts for the 350 highest-weighted features.

As shown in Table \ref{Table4:FeatureCounts}, Energy/Loudness, MFCC, Spectral Balance, and Spectral Dynamics features were over-represented among the selected features common to all classifiers, with Spectral Dynamics showing the highest enrichment (2.55). F0 (Pitch) features were also moderately over-represented. In contrast, Auditory Spectrogram, Voice Quality, and Voicing \& Temporal Structure features were under-represented relative to their original proportions. Overall, these results indicate that spectral- and energy-related characteristics play a dominant role in discriminative performance across classifiers.

\begin{table}[H]
\centering
\caption{Feature category distribution and enrichment across classifiers.}
\label{Table4:FeatureCounts}
\begin{tabular}{llll}
\hline
Category & All ($n=6923$) & All models ($n=500$) & Enrichment \\
\hline
Auditory Spectrogram & 2770 & 125 & 0.62 \\
\textbf{Energy / Loudness} & \textbf{184} & \textbf{25} & \textbf{1.88} \\
\textbf{F0 (Pitch)} & \textbf{168} & \textbf{17} & \textbf{1.40} \\
Formants & 12 & 0 & 0.00 \\
\textbf{MFCC} & \textbf{1496} & \textbf{132} & \textbf{1.22} \\
\textbf{Spectral Balance} & \textbf{1654} & \textbf{158} & \textbf{1.32} \\
\textbf{Spectral Dynamics} & \textbf{103} & \textbf{19} & \textbf{2.55} \\
Voice Quality & 310 & 17 & 0.76 \\
Voicing \& Temporal Structure & 226 & 7 & 0.43 \\
\hline
\end{tabular}
\end{table}

\subsection{Computational cost}

Table \ref{Table5:ComputationTimes} shows that EMLM was consistently the fastest method in repeated LOSO model training across both the ADReSS and extended Pitt Corpus datasets. Its model training time increased only moderately with the number of features, rising by approximately twofold when increasing from~500 to~2500 features, whereas the computation times of L-SVM and Ridge regression increased substantially, particularly with larger feature sets. This advantage is especially relevant in LOSO validation, where the model must be retrained separately for each subject. In the separate test setting, computation times were low for all methods ($\le$ 0.07 s), and thus negligible compared to the repeated LOSO training runs.

\begin{table}[H]
\centering
\caption{Computation time (seconds) for repeated model training and feature selection, reported as a median (Q1--Q3) over 100 runs. Feature selection times refer to a single 5-fold feature-weight estimation run. The full repeated procedure consisted of 100 such iterations.}
\label{Table5:ComputationTimes}
\resizebox{\textwidth}{!}{\begin{tabular}{lllll}
\hline
 & \multicolumn{3}{c}{\textbf{Model training}} & \textbf{Feature selection} \\
\cline{2-4}
Method 
& 500 features 
& 1500 features 
& 2500 features 
& (5-fold CV) \\
\hline
\multicolumn{5}{l}{\textbf{ADReSS-LOSO cross-validation}} \\
EMLM 
& 0.48 s (0.47--0.48 s) 
& 0.64 s (0.64--0.65 s) 
& 0.87 s (0.86--0.87 s) 
& 3.19 s (3.14--3.35 s) \\
L-SVM 
& 1.20 s (1.19--1.20 s) 
& 5.54 s (5.53--5.55 s) 
& 12.34 s (12.30--12.38 s) 
& 1.67 s (1.61--1.75 s) \\
Ridge 
& 11.45 s (11.45--11.45 s) 
& 37.11 s (37.10--37.17 s) 
& 55.41 s (55.39--55.43 s) 
& 0.42 s (0.36--0.44 s) \\
\hline
\multicolumn{5}{l}{\textbf{Extended Pitt Corpus-LOSO cross-validation}} \\
EMLM 
& 1.62 s (1.61--1.63 s) 
& 2.71 s (2.64--2.74 s) 
& 3.63 s (3.62--3.65 s) 
& -- \\
L-SVM 
& 4.04 s (4.04--4.05 s) 
& 20.34 s (20.30--20.39 s) 
& 47.38 s (47.17--47.59 s) 
& -- \\
Ridge 
& 36.41 s (36.38--36.42 s) 
& 99.71 s (99.45--100.21 s) 
& 167.99 s (167.93--168.27 s) 
& -- \\
\hline
\multicolumn{5}{l}{\textbf{Separate test data}} \\
EMLM 
& $<$0.01 s  
& 0.01 s (0.01--0.01 s) 
& 0.01 s (0.01--0.01 s) 
& -- \\
L-SVM 
& 0.01 s (0.01--0.01 s) 
& 0.02 s (0.01--0.02 s) 
& 0.03 s (0.02--0.03 s) 
& -- \\
Ridge 
& 0.01 s (0.01--0.01 s) 
& 0.04 s (0.04--0.04 s) 
& 0.07 s (0.07--0.07 s) 
& -- \\
\hline
\end{tabular}}
\end{table}

\section{Discussion and conclusions}
\label{sec:DiscussionAndConclusions}

This study investigated acoustic-feature-based dementia classification from spontaneous speech using classifier-based wrapper machine learning models across multiple evaluation settings. On the ADReSS dataset, all three classifiers achieved strong performance, with Ridge regression yielding the highest accuracy in LOSO cross-validation, while EMLM and L-SVM showed comparable results. Performance decreased on the independent test set and the extended Pitt Corpus, reflecting increased dataset heterogeneity. However, classification accuracy remained relatively stable across models, suggesting robustness across the evaluated datasets.

Notably, EMLM achieved competitive classification accuracy across all datasets while being clearly the most computationally efficient classifier in repeated LOSO model training. This finding is practically important in the present study because the wrapper-based feature selection procedure involved repeated model fitting to evaluate candidate feature subsets. In this setting, EMLM provides a favorable accuracy-efficiency trade-off. Although Ridge and L-SVM achieved comparable or slightly higher accuracies in some settings, EMLM maintained competitive performance with substantially lower computational cost.

When evaluated on the extended Pitt Corpus, all models maintained stable performance despite the larger and more heterogeneous sample, supporting the robustness of the proposed approach beyond the original ADReSS split, although not yet demonstrating generalization to a fully independent corpus. Compared with previously reported acoustic-only results on the same ADReSS benchmark dataset, the obtained accuracies fall within a competitive range \cite{luz2021detecting}. Importantly, this performance was achieved without deep learning architectures, which often require task-specific adaptation, careful fine-tuning, and increased computational resources \cite{baevski2020wav2vec,balagopalan2021comparing}. Although multimodal approaches combining acoustic and lexical features have reported higher accuracy \cite{cummins2020comparison,rohanian2021multi,koo2020exploiting,pompili2020inesc,edwards2020multiscale}, many of these methods depend on more complex model architectures and linguistic information, which may limit interpretability, scalability, and practical deployment in clinical settings.

In the present study, the number of selected features was determined using a permutation-based statistical cutoff, as described in Section \ref{sec:permutationTesting} and evaluated in Section \ref{sec:ResultsStatisticalEstimation}. Similar data-driven approaches for selecting acoustic feature subsets have been reported in prior work on speech-based dementia detection. For example, \cite{koo2020exploiting} used analysis-of-variance-based selection to identify 393 relevant ComParE features, whereas \cite{edwards2020multiscale} applied a two-stage selection framework combining correlation feature selection and recursive feature elimination with cross-validation, resulting in a reduced set of 54 features and 74.0\% accuracy under LOSO validation. Overall, the feature subset sizes identified in the present study fall within, or slightly above, the broad range reported in earlier work, reflecting differences in feature representations, selection strategies, and evaluation protocols.

The results indicated that spectral- and energy-related acoustic characteristics were among the most important contributors to discriminative performance across classifiers. These features capture the spectral structure of the speech signal and its temporal energy distribution, both of which are associated with articulation precision, prosodic variation, and the control of speech rhythm and vocal intensity, as commonly described in acoustic speech analysis literature \cite{kappen2024acoustic}. Clinical studies have shown that speech produced by individuals with dementia is characterized by increased pausing, slower speech rate, and irregularities in rhythm and prosody compared to healthy controls, and that these alterations are reflected in temporal, spectral, and energy-related acoustic measures \cite{saeedi2024acoustic,jafari2025diagnostic}.

In the present study, acoustic features were extracted from full recordings rather than from speech-active segments only. This design choice enabled a compact recording-level representation while retaining temporal characteristics, such as hesitation and silent pauses, that may otherwise be excluded in segment-level approaches. Nevertheless, retaining non-speech regions should be considered when interpreting the results, as some acoustic features may also reflect non-speech-related cues. In particular, energy- and loudness-related features may partly reflect the proportion and acoustic properties of silent intervals rather than vocal intensity during speech. Because Energy/Loudness features were over-represented among the highest-ranked features, future work would benefit from explicitly comparing non-speech energy distributions between AD and Non-AD groups.

Future research could benefit from increasing the dataset size by including a larger group of participants, which may improve the reliability of the results. The DementiaBank database from the TalkBank project includes several corpora, such as the Wisconsin Longitudinal Study (WLS) corpus, which is a large longitudinal dataset \cite{herd2014cohort}. This dataset contains recordings from the image description task, as well as multiple scored linguistic tasks that assess word retrieval speed (verbal fluency), short-term memory, semantic memory, and linguistic reasoning skills.

The Taukadial corpus contains a total of 507 recordings, of which 261 are in Chinese and 246 in English \cite{luz2024connected}. Recordings were collected from study participants performing three different tasks: two picture description tasks and one narrative task. The dataset would enable empirical evaluation of the cross-linguistic portability of acoustic feature-based machine learning models, such as those proposed in the present study.

A relevant future direction is to reconsider the formulation of the machine learning task as a binary classification problem, in which subjects are categorized simply as having dementia or not. This binary formulation represents a methodological simplification and does not fully reflect clinical practice, where dementia progresses through distinct stages and cognitive assessment instruments such as the MMSE, MoCA, and CERAD use predefined score thresholds to support clinical interpretation. If cognitive test scores were available alongside speech recordings, a more clinically aligned approach would involve stage-based classification. Alternatively, the task could be reformulated as a regression problem aimed at directly predicting cognitive test scores. From a screening perspective, higher sensitivity is desirable, since missed Alzheimer's disease cases are typically of greater concern than false positives. The relatively high sensitivity observed especially in the extended Pitt Corpus experiments (Section \ref{sec:classificationResults}) is consistent with this clinical priority. Overall, these findings support EMLM as a particularly suitable classifier for acoustic dementia classification, especially when repeated model training in LOSO validation makes computational efficiency an important practical consideration.

\section*{Use of AI tools declaration}
The authors declare they have not used Artificial Intelligence (AI) tools in the creation of this article.

\section*{Acknowledgments}
The first author (MN) would like to thank the Finnish Cultural Foundation for the financial support (Grant No.  30242323). The work of the second author (MvB) was supported by the Samfundet Folkhälsan and the Research Council of Finland (Grant No. 349336).

\section*{Conflict of interest}
Tommi Kärkkäinen is an editorial board member for Applied Computing and Intelligence and was not involved in the editorial review and the decision to publish this article.

\section*{Appendix 1. Highest-ranked acoustic features across eighteen feature categories based on average rankings from three models}
\label{sec:appendix1}


\begin{longtable}{lll}

\caption{The 100 acoustic features with the highest average ranks.}
\label{tab:top100features} \\

\hline
Rank & Feature & Category \\
\hline
\endfirsthead

\hline
Rank & Feature & Category \\
\hline
\endhead

\hline
\multicolumn{3}{r}{\textit{Continued on next page}} \\
\hline
\endfoot

\hline
\endlastfoot

1 & mfcc\_sma[7]\_lpc3 & MFCC \\ 
2 & mfcc\_sma\_de[2]\_peakRangeRel & MFCC \\ 
3 & pcm\_fftMag\_spectralRollOff90.0\_sma\_centroid & Spectral Roll-Off \\ 
4 & pcm\_zcr\_sma\_de\_quartile2 & Zero-Crossing Rate \\ 
5 & mfcc\_sma\_de[13]\_lpc3 & MFCC \\ 
6 & audSpec\_Rfilt\_sma\_de[4]\_minRangeRel & Auditory Spectrogram \\ 
7 & audSpec\_Rfilt\_sma\_de[4]\_peakRangeRel & Auditory Spectrogram \\ 
8 & pcm\_fftMag\_spectralHarmonicity\_sma\_centroid & Spectral Centroid \\ 
9 & mfcc\_sma[7]\_lpc2 & MFCC \\ 
10 & pcm\_fftMag\_psySharpness\_sma\_centroid & Spectral Centroid \\ 
11 & pcm\_fftMag\_spectralRollOff75.0\_sma\_centroid & Spectral Roll-Off \\ 
12 & pcm\_RMSenergy\_sma\_de\_meanSegLen & Energy \\ 
13 & audSpec\_Rfilt\_sma[11]\_upleveltime90 & Auditory Spectrogram \\ 
14 & pcm\_RMSenergy\_sma\_de\_maxSegLen & Energy \\ 
15 & pcm\_fftMag\_spectralCentroid\_sma\_centroid & Spectral Centroid \\ 
16 & mfcc\_sma\_de[13]\_lpc4 & MFCC \\ 
17 & audSpec\_Rfilt\_sma\_de[8]\_upleveltime75 & Auditory Spectrogram \\ 
18 & mfcc\_sma[6]\_lpc4 & MFCC \\ 
19 & mfcc\_sma\_de[11]\_peakRangeAbs & MFCC \\ 
20 & pcm\_fftMag\_spectralVariance\_sma\_centroid & Spectral Centroid \\ 
21 & pcm\_fftMag\_fband250-650\_sma\_centroid & Spectral Centroid \\ 
22 & jitterLocal\_sma\_de\_upleveltime25 & Jitter \\ 
23 & mfcc\_sma[11]\_peakMeanRel & MFCC \\ 
24 & jitterDDP\_sma\_lpc4 & Jitter \\ 
25 & mfcc\_sma\_de[4]\_meanSegLen & MFCC \\ 
26 & mfcc\_sma\_de[5]\_linregc2 & MFCC \\ 
27 & lspFreq\_sma\_de[4]\_quartile2 & Line Spectral Pairs (LSP) \\ 
28 & mfcc\_sma[12]\_linregc1 & MFCC \\ 
29 & pcm\_intensity\_sma\_linregc2 & Energy \\ 
30 & mfcc\_sma[10]\_max & MFCC \\ 
31 & mfcc\_sma[14]\_quartile3 & MFCC \\ 
32 & lspFreq\_sma\_de[0]\_quartile2 & Line Spectral Pairs (LSP) \\ 
33 & mfcc\_sma\_de[10]\_upleveltime25 & MFCC \\ 
34 & F0final\_sma\_centroid & F0 (Pitch) \\ 
35 & pcm\_fftMag\_spectralSlope\_sma\_centroid & Spectral Centroid \\
36 & pcm\_RMSenergy\_sma\_lpc4 & Energy \\ 
37 & audSpec\_Rfilt\_sma\_de[0]\_lpc4 & Auditory Spectrogram \\ 
38 & mfcc\_sma[7]\_lpc4 & MFCC \\ 
39 & audSpec\_Rfilt\_sma\_de[0]\_lpc3 & Auditory Spectrogram \\ 
40 & audSpec\_Rfilt\_sma[1]\_stddevFallingSlope & Auditory Spectrogram \\ 
41 & pcm\_fftMag\_spectralSlope\_sma\_lpc1 & Spectral Centroid \\ 
42 & audSpec\_Rfilt\_sma\_de[0]\_lpc2 & Auditory Spectrogram \\ 
43 & pcm\_fftMag\_spectralSlope\_sma\_lpc0 & Spectral Centroid \\ 
44 & mfcc\_sma[10]\_linregc2 & MFCC \\ 
45 & audSpec\_Rfilt\_sma[1]\_meanFallingSlope & Auditory Spectrogram \\ 
46 & audSpec\_Rfilt\_sma[10]\_upleveltime90 & Auditory Spectrogram \\ 
47 & pcm\_fftMag\_spectralEntropy\_sma\_centroid & Spectral Entropy \\ 
48 & pcm\_loudness\_sma\_linregc2 & Loudness \\ 
49 & lspFreq\_sma[2]\_minPos & Line Spectral Pairs (LSP) \\ 
50 & pcm\_zcr\_sma\_risetime & Zero-Crossing Rate \\ 
51 & logHNR\_sma\_de\_lpc3 & Harmonic-to-Noise Ratio (HNR) \\ 
52 & voicingFinalUnclipped\_sma\_lpc3 & Voicing Probability \\ 
53 & mfcc\_sma[10]\_upleveltime50 & MFCC \\ 
54 & pcm\_fftMag\_spectralCentroid\_sma\_linregc2 & Spectral Centroid \\ 
55 & mfcc\_sma[7]\_peakMeanRel & MFCC \\ 
56 & mfcc\_sma[10]\_qregc3 & MFCC \\ 
57 & pcm\_fftMag\_spectralRollOff90.0\_sma\_linregc1 & Spectral Roll-Off \\ 
58 & mfcc\_sma[8]\_lpc3 & MFCC \\ 
59 & lspFreq\_sma\_de[2]\_max & Line Spectral Pairs (LSP) \\ 
60 & mfcc\_sma\_de[3]\_lpc4 & MFCC \\ 
61 & pcm\_fftMag\_spectralFlux\_sma\_de\_minRangeRel & Spectral Flux \\ 
62 & pcm\_fftMag\_spectralKurtosis\_sma\_stddevFallingSlope & Spectral Kurtosis \\ 
63 & audSpec\_Rfilt\_sma\_de[0]\_lpc1 & Auditory Spectrogram \\ 
64 & mfcc\_sma[14]\_upleveltime25 & MFCC \\ 
65 & audSpec\_Rfilt\_sma[4]\_meanRisingSlope & Auditory Spectrogram \\ 
66 & mfcc\_sma\_de[8]\_minRangeRel & MFCC \\ 
67 & audSpec\_Rfilt\_sma\_de[5]\_skewness & Auditory Spectrogram \\ 
68 & jitterDDP\_sma\_de\_lpc0 & Jitter \\ 
69 & mfcc\_sma\_de[8]\_range & MFCC \\ 
70 & audSpec\_Rfilt\_sma\_de[4]\_upleveltime25 & Auditory Spectrogram \\ 
71 & audSpec\_Rfilt\_sma\_de[16]\_upleveltime50 & Auditory Spectrogram \\ 
72 & mfcc\_sma\_de[7]\_linregc2 & MFCC \\ 
73 & audSpec\_Rfilt\_sma\_de[1]\_lpc4 & Auditory Spectrogram \\ 
74 & mfcc\_sma[8]\_lpc4 & MFCC \\ 
75 & pcm\_RMSenergy\_sma\_lpc0 & Energy \\ 
76 & voiceProb\_sma\_maxPos & Voicing Probability \\ 
77 & audSpec\_Rfilt\_sma\_de[15]\_stddevRisingSlope & Auditory Spectrogram \\ 
78 & pcm\_RMSenergy\_sma\_de\_minSegLen & Energy \\ 
79 & mfcc\_sma\_de[1]\_peakRangeRel & MFCC \\ 
80 & audSpec\_Rfilt\_sma\_de[1]\_lpc0 & Auditory Spectrogram \\ 
81 & audSpec\_Rfilt\_sma\_de[1]\_lpc3 & Auditory Spectrogram \\ 
82 & audSpec\_Rfilt\_sma\_de[1]\_lpc2 & Auditory Spectrogram \\ 
83 & pcm\_RMSenergy\_sma\_centroid & Energy \\ 
84 & mfcc\_sma[1]\_minRangeRel & MFCC \\ 
85 & pcm\_fftMag\_spectralRollOff90.0\_sma\_lpc1 & Spectral Roll-Off \\ 
86 & audSpec\_Rfilt\_sma\_de[2]\_stddevRisingSlope & Auditory Spectrogram \\ 
87 & pcm\_fftMag\_fband1000-4000\_sma\_upleveltime50 & Spectral Centroid \\ 
88 & pcm\_fftMag\_spectralRollOff75.0\_sma\_de\_upleveltime90 & Spectral Roll-Off \\ 
89 & jitterDDP\_sma\_de\_centroid & Jitter \\ 
90 & pcm\_fftMag\_spectralSlope\_sma\_de\_peakDistStddev & Spectral Centroid \\ 
91 & mfcc\_sma\_de[11]\_lpc4 & MFCC \\ 
92 & pcm\_fftMag\_spectralHarmonicity\_sma\_peakRangeRel & Spectral Centroid \\ 
93 & audSpec\_Rfilt\_sma\_de[18]\_segLenStddev & Auditory Spectrogram \\ 
94 & lspFreq\_sma[2]\_quartile3 & Line Spectral Pairs (LSP) \\ 
95 & audSpec\_Rfilt\_sma\_de[1]\_lpc1 & Auditory Spectrogram \\ 
96 & audSpec\_Rfilt\_sma[11]\_upleveltime75 & Auditory Spectrogram \\ 
97 & audSpec\_Rfilt\_sma[5]\_minRangeRel & Auditory Spectrogram \\ 
98 & audSpec\_Rfilt\_sma\_de[0]\_lpc0 & Auditory Spectrogram \\ 
99 & pcm\_fftMag\_spectralCentroid\_sma\_qregc3 & Spectral Centroid \\ 
100 & mfcc\_sma\_de[9]\_peakRangeRel & MFCC \\ 

\end{longtable}


\section*{Appendix 2. Feature category counts among the highest-weighted features for each classifier}
\label{sec:appendix2}

\begin{table}[H]
\centering
\caption{Feature counts by category for the 350 highest-weighted features selected by EMLM.}
\begin{tabular}{llll}
\hline
Category & All ($N = $ 6923) & EMLM ($N = $ 350) & Enrichment \\
\hline
Auditory Spectrogram & 2770 & 86 & 0.61 \\
\textbf{Energy / Loudness} & \textbf{184} & \textbf{27} & \textbf{2.9} \\
\textbf{F0 (Pitch)} & \textbf{168} & \textbf{10} & \textbf{1.18} \\
Formants & 12 & 0 & 0 \\
\textbf{MFCC} & \textbf{1496} & \textbf{83} & \textbf{1.1} \\
\textbf{Spectral Balance} & \textbf{1654} & \textbf{99} & \textbf{1.18} \\
\textbf{Spectral Dynamics} & \textbf{103} & \textbf{18} & \textbf{3.46} \\
\textbf{Voice Quality} & \textbf{310} & \textbf{18} & \textbf{1.15} \\
Voicing \& Temporal Structure & 226 & 9 & 0.79 \\
\hline
\end{tabular}
\end{table}

\begin{table}[H]
\centering
\caption{Feature counts by category for the 350 highest-weighted features selected by L-SVM.}
\begin{tabular}{llll}
\hline
Category & All ($N = $ 6923) & L-SVM ($N = $ 350) & Enrichment \\
\hline
Auditory Spectrogram & 2770 & 93 & 0.66 \\
\textbf{Energy / Loudness} & \textbf{184} & \textbf{15} & \textbf{1.61} \\
\textbf{F0 (Pitch)} & \textbf{168} & \textbf{10} & \textbf{1.18} \\
Formants & 12 & 0 & 0 \\
\textbf{MFCC} & \textbf{1496} & \textbf{102} & \textbf{1.35} \\
Spectral Balance & 1654 & 90 & 1.08 \\
\textbf{Spectral Dynamics} & \textbf{103} & \textbf{11} & \textbf{2.11} \\
\textbf{Voice Quality} & \textbf{310} & \textbf{19} & \textbf{1.21} \\
Voicing \& Temporal Structure & 226 & 10 & 0.88 \\
\hline
\end{tabular}
\end{table}

\begin{table}[H]
\centering
\caption{Feature counts by category for the 350 highest-weighted features selected by Ridge.}
\begin{tabular}{llll}
\hline
Category & All ($N = $ 6923) & Ridge ($N = $ 350) & Enrichment \\
\hline
Auditory Spectrogram & 2770 & 86 & 0.61 \\
\textbf{Energy / Loudness} & \textbf{184} & \textbf{17} & \textbf{1.83} \\
\textbf{F0 (Pitch)} & \textbf{168} & \textbf{14} & \textbf{1.65} \\
Formants & 12 & 0 & 0 \\
MFCC & 1496 & 78 & 1.03 \\
\textbf{Spectral Balance} & \textbf{1654} & \textbf{130} & \textbf{1.55} \\
Spectral Dynamics & 103 & 4 & 0.77 \\
Voice Quality & 310 & 9 & 0.57 \\
Voicing \& Temporal Structure & 226 & 12 & 1.05 \\
\hline
\end{tabular}

\end{table}


\begin{thebibliography}{999}
	
	\bibitem{pasquier1999early}
 F.~Pasquier,
 Early diagnosis of dementia: neuropsychology,
 \emph{J. Neurol.}, \textbf{246} (1999), 6--15. \doilink{https://doi.org/10.1007/s004150050299}
	
	\bibitem{world2017global}
 World Health Organization,
 \emph{Global action plan on the public health response to dementia
		2017--2025}, Geneva: WHO Document Production Services, 2017.  
	
	\bibitem{wang2019cognitive}
 Y.~Wang, M.~L. Haaksma, I.~H. Ramakers, F.~R. Verhey, W.~M. van~de
	Flier, P.~Scheltens, et~al.,
 Cognitive and functional progression of dementia in two longitudinal
	studies,
 \emph{Int. J. Geriatr. Psych.}, \textbf{34}
	(2019), 1623--1632. \doilink{https://doi.org/10.1002/gps.5175}
	
	\bibitem{folstein1975mini}
 M.~F. Folstein, S.~E. Folstein, P.~R. McHugh,
 “Mini-mental state”: a practical method for grading the cognitive state of patients for the clinician,
 \emph{J. Psychiatr. Res.}, \textbf{12} (1975), 189--198. \doilink{https://doi.org/10.1016/0022-3956(75)90026-6}
	
	\bibitem{nasreddine2005montreal}
 Z.~S. Nasreddine, N.~A. Phillips, V.~B{\'e}dirian, S.~Charbonneau,
	V.~Whitehead, I.~Collin, et al.,
 The montreal cognitive assessment, moca: a brief screening tool for
	mild cognitive impairment,
 \emph{J. Amer. Geriatr. Soc.}, \textbf{53}
	(2005), 695--699. \doilink{https://doi.org/10.1111/j.1532-5415.2005.53221.x}
	
	\bibitem{welsh1994consortium}
 K.~A. Welsh, N.~Butters, R.~C. Mohs, D.~Beekly, S.~Edland,
	G.~Fillenbaum et~al.,
 The consortium to establish a registry for alzheimer's disease
	(cerad), part v: a normative study of the neuropsychological battery,
 \emph{Neurology}, \textbf{44} (1994), 609--609. \doilink{https://doi.org/10.1212/WNL.44.4.609}
	
	\bibitem{zorluoglu2015mobile}
 G.~Zorluoglu, M.~E. Kamasak, L.~Tavacioglu, P.~O. Ozanar,
 A mobile application for cognitive screening of dementia,
 \emph{Comput. Meth. Prog. Bio.}, \textbf{118}
	(2015), 252--262. \doilink{https://doi.org/10.1016/j.cmpb.2014.11.004}
	
	\bibitem{jack2018nia}
 C.~R. Jack, D.~A. Bennett, K.~Blennow, M.~C. Carrillo, B.~Dunn,
	S.~B. Haeberlein,   et~al.,
 Nia-aa research framework: toward a biological definition of
	alzheimer's disease,
 \emph{Alzh. Dement.}, \textbf{14} (2018), 535--562. \doilink{https://doi.org/10.1016/j.jalz.2018.02.018}
	
	\bibitem{hanyu2000magnetization}
 H.~Hanyu, T.~Asano, T.~Iwamoto, M.~Takasaki, H.~Shindo, K.~Abe,
 Magnetization transfer measurements of the hippocampus in patients
	with alzheimer's disease, vascular dementia, and other types of dementia,
 \emph{Amer. J. Neuroradiol.}, \textbf{21} (2000),
	1235--1242. 
	
	\bibitem{ebrahimighahnavieh2020deep}
 M.~A. Ebrahimighahnavieh, S.~Luo, R.~Chiong,
 Deep learning to detect alzheimer's disease from neuroimaging: a
	systematic literature review,
 \emph{Comput. Meth. Prog. Bio.}, \textbf{187}
	(2020), 105242. \doilink{https://doi.org/10.1016/j.cmpb.2019.105242}
	
	\bibitem{mirzaei2022machine}
 G.~Mirzaei, H.~Adeli,
 Machine learning techniques for diagnosis of alzheimer disease, mild
	cognitive disorder, and other types of dementia,
 \emph{Biomed. Signal Proces.}, \textbf{72} (2022),
	103293. \doilink{https://doi.org/10.1016/j.bspc.2021.103293}
	
	\bibitem{young2020imaging}
 P.~N. Young, M.~Estarellas, E.~Coomans, M.~Srikrishna, H.~Beaumont,
	A.~Maass,  
	et~al.,
 Imaging biomarkers in neurodegeneration: current and future
	practices,
 \emph{Alz. Res. Therapy}, \textbf{12} (2020), 49. \doilink{https://doi.org/10.1186/s13195-020-00612-7}
	
	\bibitem{reilly2011anomia}
 J.~Reilly, J.~E. Peelle, S.~M. Antonucci, M.~Grossman,
 Anomia as a marker of distinct semantic memory impairments in
	alzheimer's disease and semantic dementia,
 \emph{Neuropsychology}, \textbf{25} (2011), 413--426. \doilink{https://doi.org/10.1037/a0022738}
	
	\bibitem{ivanova}
 O.~Ivanova, I.~Mart{\'\i}nez-Nicol{\'a}s, E.~Garc{\'\i}a-Pi{\~n}uela, J.~J.~G. Meil{\'a}n,
 Defying syntactic preservation in alzheimer's disease: what type of
	impairment predicts syntactic change in dementia (if it does) and why? 
 \emph{Front. Lang. Sci.}, \textbf{2} (2023), 1199107. \doilink{https://doi.org/10.3389/flang.2023.1199107}
	
	\bibitem{fraser2015linguistic}
 K.~C. Fraser, J.~A. Meltzer and F.~Rudzicz,
 Linguistic features identify alzheimer’s disease in narrative
	speech,
 \emph{Journal of Alzheimer's Disease}, \textbf{49} (2015), 407--422. \doilink{https://doi.org/10.3233/JAD-150520}
	
	\bibitem{de2020artificial}
 S.~De~la Fuente~Garcia, C.~W. Ritchie, S.~Luz,
 Artificial intelligence, speech, and language processing approaches
	to monitoring alzheimer’s disease: a systematic review,
 \emph{Journal of Alzheimer's Disease}, \textbf{78} (2020),
	1547--1574. \doilink{https://doi.org/10.3233/JAD-200888}
	
	\bibitem{zhen2023}
 H.~Zhen, Y.~Shi, J.~J. Yang, J.~M. Vehni,
 Co-supervised learning paradigm with conditional generative
	adversarial networks for sample-efficient classification,
 \emph{Applied Computing and Intelligence}, \textbf{3} (2023), 13--26. \doilink{https://doi.org/10.3934/aci.2023002}
	
	\bibitem{alsuhaibani2025review}
 M.~Alsuhaibani, A.~Pourramezan~Fard, J.~Sun, F.~Far~Poor, P.~S.
	Pressman, M.~H. Mahoor,
 A review of machine learning approaches for non-invasive cognitive
	impairment detection,
 \emph{IEEE Access}, \textbf{13} (2025), 56355--56384. \doilink{https://doi.org/10.1109/ACCESS.2025.3555176}
	
	\bibitem{ortiz2023deep}
 D.~Ortiz-Perez, P.~Ruiz-Ponce, D.~Tom{\'a}s, J.~Garcia-Rodriguez,
	M.~F. Vizcaya-Moreno, M.~Leo,
 A deep learning-based multimodal architecture to predict signs of
	dementia,
 \emph{Neurocomputing}, \textbf{548} (2023), 126413. \doilink{https://doi.org/10.1016/j.neucom.2023.126413}
	
	\bibitem{csahin2025unlocking}
 E.~{\c{S}}AHiN, N.~N. Arslan, D.~{\"O}zdemir,
 Unlocking the black box: an in-depth review on interpretability,
	explainability, and reliability in deep learning,
 \emph{Neural Comput. Appl.}, \textbf{37} (2025),
	859--965. \doilink{https://doi.org/10.1007/s00521-024-10437-2}
	
	\bibitem{linja2023feature}
 J.~Linja, J.~H{\"a}m{\"a}l{\"a}inen, P.~Nieminen,
	T.~K{\"a}rkk{\"a}inen,
 Feature selection for distance-based regression: An umbrella review
	and a one-shot wrapper,
 \emph{Neurocomputing}, \textbf{518} (2023), 344--359. \doilink{https://doi.org/10.1016/j.neucom.2022.11.023}
	
	\bibitem{kumar2022dementia}
 M.~R. Kumar, S.~Vekkot, S.~Lalitha, D.~Gupta, V.~J. Govindraj,
	K.~Shaukat, et al.,
 Dementia detection from speech using machine learning and deep
	learning architectures,
 \emph{Sensors}, \textbf{22} (2022), 9311. \doilink{https://doi.org/10.3390/s22239311}
	
	\bibitem{luz2021alzheimer}
 S.~Luz, F.~Haider, S.~de~la Fuente~Garcia, D.~Fromm, 
	B.~MacWhinney,
 Alzheimer's dementia recognition through spontaneous speech,
 \emph{Front. Comput. Sci.}, \textbf{3} (2021), 780169. \doilink{https://doi.org/10.3389/fcomp.2021.780169} 
	
	\bibitem{cortes1995support}
 C.~Cortes, V.~Vapnik,
 Support-vector networks,
 \emph{Mach. Learn.}, \textbf{20} (1995), 273--297. \doilink{https://doi.org/10.1007/BF00994018}
	
	\bibitem{hastie2009elements}
 T.~Hastie, R.~Tibshirani, J.~Friedman,
 \textit{The elements of statistical learning: data mining, inference, and prediction}, 2 Eds., New York: Springer, 2009. \doilink{https://doi.org/10.1007/978-0-387-84858-7}
	
	\bibitem{karkkainen2019extreme}
 T.~K{\"a}rkk{\"a}inen,
 Extreme minimal learning machine: ridge regression with
	distance-based basis,
 \emph{Neurocomputing}, \textbf{342} (2019), 33--48. \doilink{https://doi.org/10.1016/j.neucom.2018.12.078}
	
	\bibitem{meilan}
 J.~J.~G. Meil{\'a}n, F.~Mart{\'\i}nez-S{\'a}nchez, J.~Carro, D.~E.
	L{\'o}pez, L.~Millian-Morell, J.~M. Arana,
 Speech in alzheimer's disease: can temporal and acoustic parameters
	discriminate dementia?
 \emph{Dement. Geriatr. Cogn.}, \textbf{37}
	(2014), 327--334. \doilink{https://doi.org/10.1159/000356726}
	
	\bibitem{martinez}
 I.~Mart{\'\i}nez-Nicol{\'a}s, T.~E. Llorente,
	F.~Mart{\'\i}nez-S{\'a}nchez, J.~J.~G. Meil{\'a}n,
 Ten years of research on automatic voice and speech analysis of
	people with alzheimer's disease and mild cognitive impairment: a systematic
	review article,
 \emph{Front. Psychol.}, \textbf{12} (2021), 620251. \doilink{https://doi.org/10.3389/fpsyg.2021.620251}
	
	\bibitem{scherer2003vocal}
 K.~R. Scherer, T.~Johnstone, G.~Klasmeyer,
 Vocal expression of emotion,
 In: \emph{Handbook of affective sciences}, 
 Oxford: Oxford University Press,  2003,
 433--456. \doilink{https://doi.org/10.1093/oso/9780195126013.003.0023}
	
	\bibitem{pistono2016pauses}
 A.~Pistono, M.~Jucla, E.~J. Barbeau, L.~Saint-Aubert, B.~Lemesle,
	B.~Calvet, et al.,
 Pauses during autobiographical discourse reflect episodic memory
	processes in early alzheimer's disease,
 \emph{Journal of Alzheimer's Disease}, \textbf{50} (2016), 687--698. \doilink{https://doi.org/10.3233/JAD-150408}
	
	\bibitem{pappagari2020using}
 R.~Pappagari, J.~Cho, L.~Moro-Vel{\'a}zquez, N.~Dehak,
 Using state of the art speaker recognition and natural language
	processing technologies to detect alzheimer's disease and assess its
	severity, \emph{Proceedings of Interspeech 2020}, 2020,
 2177--2181. \doilink{https://doi.org/10.21437/Interspeech.2020-2587}
	
	\bibitem{meghanani2021exploration}
 A.~Meghanani, C.~S. Anoop, A.~Ramakrishnan,
 An exploration of log-mel spectrogram and mfcc features for
	alzheimer’s dementia recognition from spontaneous speech,
 \emph{Proceedings of IEEE Spoken Language Technology Workshop (SLT)}, 2021,
 670--677. \doilink{https://doi.org/10.1109/SLT48900.2021.9383491}
	
	\bibitem{parlak2023voice}
 M.~M. Parlak, G.~Saylam, M.~A. Babademez, {\"O}.~B. Munis, S.~A.
	Tokg{\"o}z,
 Voice analysis results in individuals with alzheimer's disease: How
	do age and cognitive status affect voice parameters?
 \emph{Brain Behav.}, \textbf{13} (2023), e3271. \doilink{https://doi.org/10.1002/brb3.3271}
	
	\bibitem{eyben2015geneva}
 F.~Eyben, K.~R. Scherer, B.~W. Schuller, J.~Sundberg, E.~Andr{\'e},
	C.~Busso,  et~al.,
 The geneva minimalistic acoustic parameter set (gemaps) for voice
	research and affective computing,
 \emph{IEEE Trans. Affect. Comput.}, \textbf{7} (2015),
	190--202. \doilink{https://doi.org/10.1109/TAFFC.2015.2457417}
	
	\bibitem{eyben2015realtime}
 F.~Eyben,
 \emph{Real-time speech and music classification by large audio
		feature space extraction},
 Cham: Springer, 2015. \doilink{https://doi.org/10.1007/978-3-319-27299-3}
	
	\bibitem{degottex}
 G.~Degottex, J.~Kane, T.~Drugman, T.~Raitio,  S.~Scherer,
 Covarep—a collaborative voice analysis repository for speech
	technologies,
 \emph{Proceedings of IEEE International Conference on Acoustics, Speech and Signal Processing}, 2014,
 960--964. \doilink{https://doi.org/10.1109/ICASSP.2014.6853739}
	
	\bibitem{snyder2018x}
 D.~Snyder, D.~Garcia-Romero, G.~Sell, D.~Povey,  S.~Khudanpur,
 X-vectors: robust dnn embeddings for speaker recognition,
 \emph{Proceedings of  IEEE international conference on acoustics, speech and
		signal processing (ICASSP)}, 2018,
 5329--5333. \doilink{https://doi.org/10.1109/ICASSP.2018.8461375}
	
	\bibitem{gemmeke2017audio}
 J.~F. Gemmeke, D.~P. Ellis, D.~Freedman, A.~Jansen, W.~Lawrence,
	R.~C. Moore, et al.,
 Audio set: an ontology and human-labeled dataset for audio events,
  \emph{Proceedings of  IEEE international conference on acoustics, speech and
		signal processing (ICASSP)}, 2017, 
 776--780. \doilink{https://doi.org/10.1109/ICASSP.2017.7952261}
	
	\bibitem{cummins2020comparison}
 N.~Cummins, Y.~Pan, Z.~Ren, J.~Fritsch, V.~S. Nallanthighal,
	H.~Christensen,  
	et~al.,
 A comparison of acoustic and linguistics methodologies for
	alzheimer’s dementia recognition,  \emph{Proceedings of Interspeech 2020},
 2020,
 2182--2186. \doilink{https://doi.org/10.21437/Interspeech.2020-2635}
	
	\bibitem{baevski2020wav2vec}
 A.~Baevski, Y.~Zhou, A.~Mohamed, M.~Auli,
 Wav2vec 2.0: A framework for self-supervised learning of speech
	representations,
 \emph{ Proceedings of the 34th International Conference on Neural Information Processing Systems},  
	2020, 12449--12460. 
	
	\bibitem{balagopalan2021comparing}
 A.~Balagopalan, B.~Eyre, J.~Robin, F.~Rudzicz, J.~Novikova,
 Comparing pre-trained and feature-based models for prediction of
	{A}lzheimer's disease based on speech,
 \emph{Front. Aging Neurosci.}, \textbf{13} (2021), 635945. \doilink{https://doi.org/10.3389/fnagi.2021.635945}
	
	\bibitem{dehak2010front}
 N.~Dehak, P.~J. Kenny, R.~Dehak, P.~Dumouchel, P.~Ouellet,
 Front-end factor analysis for speaker verification,
 \emph{IEEE Trans. Audio Speech.},
	\textbf{19} (2010), 788--798. \doilink{https://doi.org/10.1109/TASL.2010.2064307}
	
	\bibitem{schmitt2017openxbow}
 M.~Schmittm B.~Schuller,
 openXBOW---Introducing the Passau open-source crossmodal Bag-of-Words toolkit,
 \emph{J. Mach. Learn. Res.}, \textbf{18} (2017),
	1--5. 
	
	\bibitem{syed2020automated}
 M.~S.~S. Syed, Z.~S. Syed, M.~Lech, E.~Pirogova,
 Automated screening for alzheimer's dementia through spontaneous
	speech,
 \emph{Proceedings of Interspeech 2020}, 2020, 
 2222--2226. \doilink{https://doi.org/10.21437/Interspeech.2020-3158}
	
	\bibitem{haider2019assessment}
 F.~Haider, S.~De~La~Fuente, S.~Luz,
 An assessment of paralinguistic acoustic features for detection of
	alzheimer's dementia in spontaneous speech,
 \emph{IEEE J.-STSP},
	\textbf{14} (2019), 272--281. \doilink{https://doi.org/10.1109/JSTSP.2019.2955022}
	
	\bibitem{niemela2024classification}
 M.~Niemel{\"a}, M.~von Bonsdorff, S.~{\"A}yr{\"a}m{\"o}, 
	T.~K{\"a}rkk{\"a}inen,
 Classification of dementia from spoken speech using feature selection
	and the bag of acoustic words model,
 \emph{Applied Computing and Intelligence}, \textbf{4} (2024), 45--65. \doilink{https://doi.org/10.3934/aci.2024004} 
	
	\bibitem{martinc2020tackling}
 M.~Martinc, S.~Pollak,
 Tackling the adress challenge: a multimodal approach to the automated
	recognition of alzheimer's dementia,
 \emph{Proceedings of Interspeech 2020}, 2020,
 2157--2161. \doilink{https://doi.org/10.21437/Interspeech.2020-2202} 
	
	\bibitem{rohanian2021multi}
 M.~Rohanian, J.~Hough, M.~Purver,
 Multi-modal fusion with gating using audio, lexical and disfluency
	features for alzheimer's dementia recognition from spontaneous speech,
  {\it Proceedings of Interspeech 2020}, 2020, 2187--2191. \doilink{https://doi.org/10.21437/Interspeech.2020-2721
  }
	
	\bibitem{becker1994natural}
 J.~T. Becker, F.~Boiler, O.~L. Lopez, J.~Saxton,  K.~L. McGonigle,
 The natural history of alzheimer's disease: description of study
	cohort and accuracy of diagnosis,
 \emph{Arch. Neurol.}, \textbf{51} (1994), 585--594. \doilink{https://doi.org/10.1001/ARCHNEUR.1994.00540180063015}
	
	\bibitem{fong2019factor}
 M.~W. Fong, R.~Van~Patten, R.~P. Fucetola,
 The factor structure of the boston diagnostic aphasia examination,
 \emph{J. Int. Neuropsych. Soc.},
	\textbf{25} (2019), 772--776. \doilink{https://doi.org/10.1017/S1355617719000237}
	
	\bibitem{macwhinney2014childes}
 B.~MacWhinney,
 \emph{The CHILDES project: Tools for analyzing talk, volume I:
		transcription format and programs}, 
 New York: Psychology Press, 2014. \doilink{https://doi.org/10.4324/9781315805672}
	
	\bibitem{ebu2011loudness}
 \textit{R.~EBU-Recommendation,
 Loudness normalisation and permitted maximum level of audio signals},
 European Broadcasting Union,
 2011.  Available from:  \url{https://tech.ebu.ch/publications/r128}. 
	
	\bibitem{itutp835}
 
 \emph{{International Telecommunication Union}, Subjective test methodology for evaluating speech communication
		systems that include noise suppression algorithm},
ITU-T Publications, 2003. Available from:  \url{https://www.itu.int/rec/T-REC-P.835/en}.
	
	\bibitem{eyben}
 F.~Eyben, M.~W{\"o}llmer, B.~Schuller,
 Opensmile: the munich versatile and fast open-source audio feature
	extractor,
  \emph{Proceedings of the 18th ACM International Conference on
		Multimedia}, 2010,
 1459--1462. \doilink{https://doi.org/10.1145/1873951.1874246}
		
	\bibitem{hamalainen2020minimal}
 J.~H{\"a}m{\"a}l{\"a}inen, A.~S.~C. Alencar, T.~Kärkkäinen,
	C.~L.~C. Mattos, H.~S. Amauri, J.~P.~P. Gomes,
 Minimal learning machine: theoretical results and clustering-based
	reference point selection,
 \emph{J. Mach. Learn. Res.}, \textbf{21} (2020),
	1--29.  
	
	\bibitem{karkkainen2019model}
 T.~K{\"a}rkk{\"a}inen,
 Model selection for extreme minimal learning machine using sampling,
 \emph{Proceedings of the 27th European Symposium on
		Artificial Neural Networks, Computational Intelligence and Machine Learning},
	2019,
 391--396. 
	
	\bibitem{karkkainen2005computation}
 T.~K{\"a}rkk{\"a}inen, S.~{\"A}yr{\"a}m{\"o},
 On computation of spatial median for robust data mining,
 \emph{Proceedings of Evolutionary and Deterministic Methods for Design,
		Optimization and Control with Applications to Industrial and Societal
		Problems}, 2005,
 1--14.  
	
	\bibitem{hastie2014glmnet}
\textit{J.  Qian, T. Hastie, J. Friedman, R. Tibshirani, N. Simon, 
 Glmnet for matlab}, jemdoc, 2014. Available from:  
 \url{https://hastie.su.domains/glmnet_matlab/}.
	
	\bibitem{breiman2001random}
 L.~Breiman,
 Random forests,
 \emph{Mach. Learn.}, \textbf{45} (2001), 5--32. \doilink{https://doi.org/10.1023/A:1010933404324}
	
	\bibitem{gibbons2021nonparametric}
 J.~D. Gibbons, S.~Chakraborti,
 \emph{Nonparametric statistical inference},
 New York: Chapman and Hall/CRC, 2010. \doilink{https://doi.org/10.1201/9781439896129}
	
	\bibitem{jauhiainen2021new}
 S.~Jauhiainen, J. P. Kauppi, M.~Lepp{\"a}nen, K.~Pasanen,
	J.~Parkkari, T.~Vasankari, et al.,
 New machine learning approach for detection of injury risk factors in
	young team sport athletes,
 \emph{Int. J. Sports Med.}, \textbf{42} (2021),
	175--182. \doilink{https://doi.org/10.1055/a-1231-5304}
	
	\bibitem{saarela2021comparison}
 M.~Saarela, S.~Jauhiainen,
 Comparison of feature importance measures as explanations for
	classification models,
 \emph{SN Appl. Sci.}, \textbf{3} (2021), 272. \doilink{https://doi.org/10.1007/s42452-021-04148-9}
	
	\bibitem{luz2021detecting}
 S.~Luz, F.~Haider, S.~De~la Fuente, D.~Fromm,  B.~MacWhinney,
 Detecting cognitive decline using speech only: the adresso challenge,
  arXiv: 2104.09356. \doilink{https://doi.org/10.48550/arXiv.2104.09356}
	
	\bibitem{koo2020exploiting}
 J.~Koo, J.~H. Lee, J.~Pyo, Y.~Jo, K.~Lee,
 Exploiting multi-modal features from pre-trained networks for
	alzheimer's dementia recognition,
  arXiv: 2009.04070. \doilink{https://doi.org/10.48550/arXiv.2009.04070}
	
	\bibitem{pompili2020inesc}
 A.~Pompili, T.~Rolland, A.~Abad,
 The inesc-id multi-modal system for the adress 2020 challenge, arXiv: 2005.14646.  \doilink{https://doi.org/10.48550/arXiv.2005.14646}
	
	\bibitem{edwards2020multiscale}
 E.~Edwards, C.~Dognin, B.~Bollepalli, M.~K. Singh,  V.~Analytics,
 Multiscale system for alzheimer's dementia recognition through
	spontaneous speech,
 \emph{Proceedings of Interspeech 2020}, 2020,
 2197--2201. \doilink{https://doi.org/10.21437/Interspeech.2020-2781} 
	
	\bibitem{kappen2024acoustic}
 M.~Kappen, G.~Vanhollebeke, J.~Van Der~Donckt, S.~Van~Hoecke, 
	M. A. Vanderhasselt,
 Acoustic and prosodic speech features reflect physiological stress
	but not isolated negative affect: a multi-paradigm study on psychosocial
	stressors,
 \emph{Sci. Rep.}, \textbf{14} (2024), 5515. \doilink{https://doi.org/10.1038/s41598-024-55550-3}
	
	\bibitem{saeedi2024acoustic}
 S.~Saeedi, S.~Hetjens, M.~Grimm, B.~B.~v. Latoszek,
 Acoustic speech analysis in alzheimer's disease: A systematic review
	and meta-analysis,
 \emph{The Journal of Prevention of Alzheimer's Disease}, \textbf{11}
	(2024), 1789--1797. \doilink{https://doi.org/10.14283/jpad.2024.132}
	
	\bibitem{jafari2025diagnostic}
 Z.~Jafari, M.~K. Andrew, K.~J. Rockwood,
 Diagnostic utility of speech-based biomarkers in mild cognitive
	impairment: a systematic review and meta-analysis,
 \emph{Age Ageing}, \textbf{54} (2025), afaf316. \doilink{https://doi.org/10.1093/ageing/afaf316}
	
	\bibitem{herd2014cohort}
 P.~Herd, D.~Carr, C.~Roan,
 Cohort profile: Wisconsin longitudinal study (wls),
 \emph{Int. J. Epidemiol.}, \textbf{43} (2014),
	34--41. \doilink{https://doi.org/10.1093/ije/dys194}
	
	\bibitem{luz2024connected}
 S.~Luz, S.~D. L.~F. Garcia, F.~Haider, D.~Fromm, B.~MacWhinney,
	A.~Lanzi, et al.,
 Connected speech-based cognitive assessment in chinese and english,
 arXiv: 2406.10272. \doilink{https://doi.org/10.48550/arXiv.2406.10272}
	
\end{thebibliography}
\end{document}